\documentclass[11pt,a4paper]{article}

\usepackage[utf8x]{inputenc}
\usepackage[T1]{fontenc}
\usepackage{color}
\usepackage{amssymb}
\usepackage{amsmath,amsfonts}
\usepackage{graphicx}
\usepackage{mathtools}
\usepackage{ragged2e}
\usepackage{psfrag}
\usepackage{relsize}
\usepackage{upgreek}
\usepackage{booktabs}

\usepackage{bm}
\usepackage{mathrsfs}
\usepackage{graphicx}
\usepackage{verbatim} 
\usepackage[english]{babel}
\usepackage[hypertexnames=false]{hyperref} 
\hypersetup{
citecolor=red,
colorlinks=true,
filecolor=red,
linkcolor=blue,
linktocpage=true,
urlcolor=blue
} 
\usepackage[titletoc,toc,title]{appendix}
\numberwithin{equation}{section}
\usepackage{cleveref}
\usepackage{cite}
\usepackage{epsfig}
\usepackage{float}
\usepackage{enumitem}
\usepackage[font={footnotesize,it}]{caption}
\usepackage{cases}
\usepackage{tikz,empheq}

\newcommand{\spll}{{/\kern-0.2em/}}
\newcommand\trick[1]{}
\newcommand{\be}{\begin{equation}} 
\newcommand{\ee}{\end{equation}}
\newcommand{\eq}[1]{(\ref{#1})}
\newcommand{\bit}{\begin{itemize}}  \newcommand{\eit}{\end{itemize}}
\newcommand{\ben}{\begin{enumerate}}  \newcommand{\een}{\end{enumerate}}

\newcommand{\bra}[1]{\langle #1|}
\newcommand{\ket}[1]{|#1 \rangle}

\newcommand{\rf}[1]{(\ref{#1})}

\def\bd{\begin{document}}
\def\ed{\end{document}}
\def\bea{\begin{eqnarray}}
\def\eea{\end{eqnarray}}

\def\la{\langle}
\def\ra{\rangle}
\newcommand{\mpl}{\ell_{\rm P}}

\def\npb#1#2#3{Nucl. Phys. {\bf{B#1}} #3 (#2)}
\def\plb#1#2#3{Phys. Lett. {\bf{#1B}} #3 (#2)}
\def\prl#1#2#3{Phys. Rev. Lett. {\bf{#1}} #3 (#2)}
\def\prd#1#2#3{Phys. Rev. {D bf{#1}} #3 (#2)}
\def\cmp#1#2#3{Comm. Math. Phys. {\bf{#1}} #3 (#2)}
\def\cqg#1#2#3{Class. Quantum Grav. {\bf{#1}} #3 (#2)}
\def\nppsa#1#2#3{Nucl. Phys. B (Proc. Suppl.) {\bf{#1A}}#3 (#2)}
\def\ap#1#2#3{Ann. of Phys. {\bf{#1}} #3 (#2)}
\def\ijmp#1#2#3{Int. J. Mod. Phys. {\bf{A#1}} #3 (#2)}
\def\rmp#1#2#3{Rev. Mod. Phys. {\bf{#1}} #3 (#2)}
\def\mpla#1#2#3{Mod. Phys. Lett. {\bf A#1} #3 (#2)}
\def\jhep#1#2#3{J. High Energy Phys. {\bf #1} #3 (#2)}
\def\atmp#1#2#3{Adv. Theor. Math. Phys. {\bf #1} #3 (#2)}

\def\sst{\scriptscriptstyle}
\def\thetabar{\bar\theta}
\def\Tr{{\rm Tr}}
\def\dd{\mathrm d}
\def\XN{X^{(N)}}
\def\XNp{X^{(N')}}
\def\An{A^{(n)}} 
\def \lp{l_P}

\newcommand{\MP}{M_{\!P}}
\newcommand{\Mzero}{M_0}
\newcommand{\HH}{\mathrm H}
\newcommand{\EE}{\mathrm E}
\newcommand{\one}{\mathbf 1}
\newcommand{\vect}[1]{\bm{#1}}

\def\a{\alpha}      \def\da{{\dot\alpha}}  \def\dA{{\dot A}}
\def\b{\beta}       \def\db{{\dot\beta}}
\def\g{\gamma}  \def\G{\Gamma}  \def\dc{{\dot\gamma}}
\def\d{\delta}  \def\D{\Delta}  \def\ddt{\dot\delta}
\def\e{\epsilon}
\def\ve{\varepsilon}
\def\uve{\upvarepsilon}
\def\f{\phi}    \def\F{\Phi}    \def\vvf{\f}
\def\vphi{\varphi}
\def\h{\eta}
\def\k{\kappa}
\def\l{\lambda} \def\L{\Lambda}
\def\m{\mu} \def\n{\nu}
\def\o{\omega}\def\O{\Omega}
\def\p{\pi} \def\P{\Pi}
\def\r{\rho}
\def\s{\sigma}  \def\S{\Sigma}
\def\t{\tau}
\def\th{\theta} \def\Th{\Theta} \def\vth{\vartheta}
\def\X{\Xeta}
\def\z{\zeta}

\def\na{\nabla}
\def\cA{{\cal A}} \def\cB{{\cal B}} \def\cC{{\cal C}}
\def\cD{{\cal D}} \def\cE{{\cal E}} \def\cF{{\cal F}}
\def\cG{{\cal G}} \def\cH{{\cal H}} \def\cI{{\cal I}}
\def\cJ{{\mathscr J}} \def\cK{{\cal K}} \def\cL{{\cal L}}
\def\cM{{\cal M}} \def\cN{{\cal N}} \def\cO{{\cal O}}
\def\cP{{\cal P}} \def\cQ{{\cal Q}} \def\cR{{\cal R}}
\def\cS{{\cal S}} \def\cT{{\cal T}} \def\cU{{\cal U}}
\def\cV{{\cal V}} \def\cW{{\cal W}} \def\cX{{\cal X}}
\def\cY{{\cal Y}} \def\cZ{{\cal Z}}
\def\ccb{{\cal b}}
\def\ct{{\cal t}}

\def\ua{\underline{\alpha}}
\def\uc{\underline{\phantom{\alpha}}\!\!\!\gamma}
\def\um{\underline{\mu}}
\def\ud{\underline\delta}
\def\ue{\underline\epsilon}
\def\una{\underline a}\def\unA{\underline A}
\def\unb{\underline b}\def\unB{\underline B}
\def\unc{\underline c}\def\unC{\underline C}
\def\und{\underline d}\def\unD{\underline D}
\def\une{\underline e}\def\unE{\underline E}
\def\unf{\underline{\phantom{e}}\!\!\!\! f}\def\unF{\underline F}
\def\unm{\underline m}\def\unM{{\underline M}}
\def\unn{\underline n}\def\unN{{\underline N}}
\def\unp{\underline{\phantom{a}}\!\!\! p}\def\unP{\underline P}
\def\unq{\underline{\phantom{a}}\!\!\! q}
\def\unQ{\underline{\phantom{A}}\!\!\!\! Q}
\def\unH{\underline{H}}

\def\As {{A \hspace{-6.4pt} \slash}\;}
\def\bs {{b \hspace{-6.4pt} \slash}\;}
\def\Ds {{D \hspace{-6.4pt} \slash}\;}
\def\Gts {{\Gt \hspace{-6.4pt} \slash}\;}
\def\ds {{\del \hspace{-6.4pt} \slash}\;}
\def\ss {{\s \hspace{-6.4pt} \slash}\;}
\def\ks {{ k \hspace{-6.4pt} \slash}\;}
\def\ps {{p \hspace{-6.4pt} \slash}\;}
\def\xs {{x \hspace{-6.4pt} \slash}\;}
\def\pas {{{p_1} \hspace{-6.4pt} \slash}\;}
\def\pbs {{{p_2} \hspace{-6.4pt} \slash}\;}
\def\cFs {{{\cal F} \hspace{-6.4pt} \slash}\;}
\def\Dss {{D \hspace{-7.5pt} \slash}\;}
\def\dss {{\del \hspace{-7.0pt} \slash}\;}

\def\Ah{{\hat{A}}}
\def\Bh{{\hat{B}}}
\def\Dh{{\hat{D}}}
\def\Gh{{\hat{G}}}
\def\Fh{{\hat{F}}}
\def\Ih{{\hat{I}}}
\def\Jh{{\hat{J}}}
\def\Kh{{\hat{K}}}
\def\Lh{{\hat{L}}}
\def\Ph{{\hat{P}}}
\def\Rh{{\hat{R}}}
\def\Vh{{\hat{V}}}
\def\Xh{{\hat{X}}}
\def\Yh{{\hat{Y}}}

\def\hn{{\hat{n}}}

\def\ah{{\hat{\a}}}
\def\bh{{\hat{\b}}}
\def\gh{{\hat{\g}}}
\def\dh{{\hat{\d}}}
\def\lh{{\hat{\l}}}
\def\rh{{\hat{\r}}}
\def\oh{{\hat{\o}}}
\def\hh{\hat{h}}
\def\uh{\hat{u}}
\def\xh{\hat{x}}
\def\yh{\hat{y}}
\def\ph{\hat{p}}
\def\xih{\hat{\xi}}
\def\chih{\hat{\chi}}
\def\Psih{\hat{\Psi}}
\def\phih{\hat{\phi}}

\def\psit{\tilde{\psi}}
\def\Psit{\tilde{\Psi}}
\def\Psibt{\tilde{\bar{Psi}}}

\def\lambdat{\tilde {\lambda}}
\def\st{\tilde{\sigma}}

\def\delt{\tilde{\delta}}
\def\Phit{\tilde{\Phi}}
\def\Phitb{\overline{\tilde{Phi}}}
\def\tht{\tilde{\th}}
\def\lt{\tilde{\l}}
\def\chit{\tilde{\chi}}
\def\phit{\tilde{\phi}}

\def\At{\tilde{A}}
\def\Bt{\tilde{B}}
\def\Ct{\tilde{C}}
\def\Dt{\tilde{D}}
\def\Et{\tilde{E}}
\def\Ft{\tilde{F}}
\def\Gt{\tilde{G}}
\def\Ht{\tilde{H}}
\def\It{\tilde{I}}
\def\Jt{\tilde{J}}
\def\Pt{\tilde{P}}
\def\Ot{\tilde{O}}
\def\Mt{\tilde{M }}
\def\Nt{\tilde{N}}
\def\St{\tilde{S}}
\def\Vt{\tilde{V}}
\def\Xt{\tilde{X}}
\def\at{\tilde{a}}
\def\ct{\tilde{c}}
\def\dt{\tilde{d}}
\def\htt{\tilde{h}}
\def\ft{\tilde{f}}
\def\gt{\tilde{\gamma}}
\def\pt{\tilde{p}}
\def\qt{\tilde{q}}
\def\rt{\tilde{r}}
\def\tt{{\tilde{t}}}
\def\nt{\tilde{n}}
\def\ut{\tilde{u}}
\def\wt{\tilde{w}}
\def\zt{\tilde{z}}
\def\xt{\tilde{x}}
\def\yt{\tilde{y}}
\def\Psit{\tilde{\Psi}}
\def\phit{\tilde{\phi}}
\def\tD{\tilde{\D}}

\def\eb{\bar{\epsilon}}
\def\delb{\bar{\partial}}
\def\thb{\bar{\theta}}
\def\mub{\bar{\mu}}
\def\lamb{\bar{\l}}
\def\psib{\bar{\psi}}
\def\sb{\bar{\sigma}}
\def\xib{\bar{\xi}}
\def\chib{\bar{\chi}}

\def\Psib{\bar{\Psi}}
\def\Phib{\bar{\Phi}}
\def\Lamb{\bar{\Lambda}}
\def\Sb{{\overline \Sigma}}
\def\hb{\bar{h}}
\def\qb{\bar{q}}
\def\wb{\bar{w}}
\def\ub{\bar{u}}
\def\zb{{\bar{z}}}
\def\Hb{\bar{H}}
\def\Qb{{\bar Q}}
\def\ob{\overline{\omega}}

\def\Ab{{\overline A}} \def\Bb{{\overline B}}
\def\Db{{\overline D}} \def\Eb{{\overline E}} \def\Fb{{\overline F}}
\def\Gb{{\overline G}}
\def\Ib{{\overline I}}
\def\Jb{{\overline J}} \def\Kb{{\overline K}} \def\Lb{{\overline L}}
\def\Mb{{\overline M}} \def\Nb{{\overline N}} \def\Ob{{\overline O}}
\def\Pb{{\overline P}}  \def\Rb{{\overline R}}
 \def\Tb{{\overline T}} \def\Ub{{\overline U}}
\def\Vb{{\overline V}} \def\Wb{{\overline W}} \def\Xb{{\overline X}}
\def\Yb{{\overline Y}} \def\Zb{{\overline Z}}

\def\fb{{\overline f}}
\def\gb{{\overline g}}
\def\nb{{\overline n}}
\def\mb{{\overline m}}
\def\lb{{\overline l}}
\def\yb{{\overline y}}

\def\ldel{{\overleftarrow{\del}}}
\def\rdel{{\overrightarrow{\del}}}
\def\ldeldel{{\overleftarrow{\del^2}}}
\def\rdeldel{{\overrightarrow{\del^2}}}
\def\ldelb{{\overleftarrow{\bar{\del}}}}
\def\rdelb{{\overrightarrow{\bar{\del}}}}
\def\ba{{\bf a}}
\def\bk{{\bf k}}
\def\bl{{\bf l}}
\def\bp{{\bf p}}
\def\bq{{\bf q}}

\def\br{{\bf r}}
\def\bt{{\bf t}}
\def\bu{{\bf u}}
\def\bv{{\bf v}}
\def\bx{{\bf x}}
\def\by{{\bf y}}
\def\bA{{\bf A}}
\def\bR{{\bf R}}
\def\bV{{\bf V}}

\def\bz{{\boldsymbol{\zeta}}}

\def\bone{{\bf 1}}

\def\va{{\vec a}}
\def\vj{{\vec j}}
\def\vk{{\vec k}}
\def\vp{{\vec p}}
\def\vq{{\vec q}}
\def\vx{{\vec x}}
\def\vy{{\vec y}}
\def\vu{{\vec u}}
\def\vv{{\vec v}}
\def \vB{{\vec B}}
\def \vE{{\vec E}}
\def \vH{{\vec H}}
\def \vg{{\vec g}}

\def\vs{{\vec \sigma}}
\def\vtau{{\vec \tau}}

\newcommand{\ov}[1]{\overrightarrow{#1}}

\def\frA{\mathfrak{A}}
\def\frB{\mathfrak{B}}
\def\frC{\mathfrak{C}}
\def\frD{\mathfrak{D}}
\def\frE{\mathfrak{E}}
\def\frF{\mathfrak{F}}
\def\frG{\mathfrak{G}}
\def\frH{\mathfrak{H}}
\def\frM{\mathfrak{M}}
\def\frN{\mathfrak{N}}
\def\frR{\mathfrak{R}}
\def\frW{\mathfrak{W}}

\def\fra{\mathfrak{a}}
\def\frb{\mathfrak{b}}
\def\frf{\mathfrak{f}}
\def\frg{\mathfrak{g}}
\def\frh{\mathfrak{h}}
\def\frl{\mathfrak{l}}
\def\frs{\mathfrak{s}}
\def\fri{\mathfrak{i}}
\def\frj{\mathfrak{j}}

\def\ma{\mathfrak{a}}
\def\mb{\mathfrak{b}}
\def\mg{\mathfrak{g}}
\def\mh{\mathfrak{h}}
\def\mB{\mathfrak{B}}
\def\mR{\mathfrak{R}}
\def\mN{\mathfrak{N}}

\def\Xk{X^{(k)}}
\def\Xkp{X^{(k')}}
\def\psik{\psi^{(k)}}
\def\psikp{\psi^{(k')}}
\def\Nk{N_k}
\def\Nkp{N_{k'}}

\newcommand{\nn}{{\nonumber}}

\def\d{\delta}\def\D{\Delta}

\def\pa{\partial} \def\del{\partial}
\def\xx{\times}
\def\uno{\mbox{1 \kern-.59em {\rm l}}}

\def\trp{^{\top}}
\def\inv{^{-1}}
\def\dag{\dagger}
\def\pr{^{\prime}}

\def\rar{\rightarrow}
\def\lar{\leftarrow}
\def\lrar{\leftrightarrow}

\newcommand{\0}{\,\!}      %this is just NOTHING!
\def\im{\imath}
\def\jm{\jmath}

\newcommand{\tr}{\mbox{tr}}
\newcommand{\slsh}[1]{/ \!\!\!\! #1}

\newcommand{\1}{\mbox{1}\hspace{-0.25em}\mbox{l}}

\def\vac{|0\rangle}
\def\lvac{\langle 0|}

\def\hlf{\frac{1}{2}}
\def\ove#1{\frac{1}{#1}}
\newcommand{\hot}[1]{\frac{#1}{2}}

\def\Box{\square}
\def\CC {\mathbb{C}}
\def\FF {\mathbb{F}}
\def\RR{\mathbb{R}}
\def\NN{\mathbb{N}}
\def\ZZ{\mathbb{Z}}
\def\bb#1{{\bf #1}}
\def\bcomment#1{}
\def\bfhat#1{{\bf \hat{#1}}}
\def\VEV#1{\left\langle #1\right\rangle}

\newcommand{\ex}[1]{{\rm e}^{#1}} \def\ii{{\rm i}}

\newcommand{\lrbrk}[1]{\left(#1\right)}
\newcommand{\lrsbrk}[1]{\left[#1\right]}
\newcommand{\sfrac}[2]{{\textstyle\frac{#1}{#2}}}

\def\stw{{\sqrt{2}}}

\def\rf {{\rm f}}
\def\ri {{\rm i}}
\def\rj {{\rm j}}
\def\rn {{\rm n}}
\def\rk {{\rm k}}
\def\rl {{\rm l}}
\def\rr {{\rm r}}

\def\rQ {{\scriptscriptstyle \rm \cQ}}
\def\rR {{\scriptscriptstyle \rm \cR}}

\def\cQb{{\cal \Qb}}
\def\cRb{{\cal \Rb}}
\def\cWb{{\cal \Wb}}

\def\fd {{\rm N}}
\def\afd {{\overline{\rm N}}}

\def \II {I\hspace{-.1em}I\hspace{.1em}}
\def \IIA {\mbox{\II A\hspace{.2em}}}
\def \IIB {\mbox{\II B\hspace{.2em}}}
\def \gs {g^s}
\def \ls {\lambda^s}

\def \I {{\cal I}}
\def \qs {q\hspace{-.53em}/\hspace{.15em}}
\def \ks {k\hspace{-.53em}/\hspace{.15em}}
\def \YM {{\mbox{\tiny YM}}}
\def \gym {g_{\YM}}

\def \Lc {\L_c}
\def\IR{\relax{\rm I\kern-.18em R}}
\def \id {{\bf 1}}

\def\cci{\ell}
\def\ccj{\ell'}

\def\bbq{\pmb{q}}
\def\bom{\pmb{\o}}
\def\bJ{\pmb{J}}
\def\bM{\pmb{M}}
\def\bB{\pmb{B}}
\def\bn{\pmb{n}}
\def\bE{\pmb{E}}

\newcommand{\rrr}[1]{\vskip 0.2cm \noindent{\bf #1} ---}

\newcommand{\KK}{\mathcal K}

\long\def\symbolfootnote[#1]#2{\begingroup%
\def\thefootnote{\fnsymbol{footnote}}\footnote[#1]{#2}\endgroup}
\long\def\RemarkBox#1{\begin{flushleft}\fbox{\begin{minipage}
{17.5cm}{\bf Remark:} ~#1\end{minipage}}\end{flushleft}}
\newcommand{\nthu}{{\it Department of Physics, National Tsing-Hua University,
    Hsinchu 30013, Taiwan}}

\newcommand{\ctc}{{\it
Center for Theory-Computation-Data Science Research, 
National Tsing-Hua University, Hsinchu 30013, Taiwan}}

\newcommand{\ncts}{{\it Physics Division,
    National Center for Theoretical Sciences, Taipei 10617, Taiwan}}

\begin{document}
\begin{center}
\vspace{20pt}
  
\thispagestyle{empty}
              {\Large \bf  Quantum Horizon and
                Quantum Membrane Paradigm \\from\\
                Black Hole Quantum Mechanics}
               
\vspace{25pt}

Chong-Sun Chu

\vspace{0.2cm}              

\vspace{5pt}\nthu\\
\vspace{5pt}\ctc\\
\vspace{5pt}\ncts

\vspace{1cm}

\begin{abstract}
We develop a microscopic quantum membrane paradigm from the matrix
quantum mechanics of black holes \cite{Chu:2024qil}.  It was proposed
that a quantum black
hole is described by a fuzzy sphere together with a half-filled Fermi
sea of horizon partons. We show that the topology of the fuzzy sphere
induces a Berry monopole, providing a microscopic origin for the
monopole appearing in the tunneling description of the decay of 
quantum black hole.  Because the partons couple to the fuzzy sphere as
fundamental degrees of freedom, they are sensitive to this monopole
and form Lowest-Landau-Level (LLL) states rather than ordinary
propagating two-dimensional fermions. Their guiding-center dynamics
generates Ohmic, Hall, and polarization currents on the horizon.  To
couple these currents to an external electromagnetic field, we
introduce a two-block matrix configuration comprising black-hole,
environmental, and bifundamental link sectors. Off-diagonal link modes
become tachyonic near the fuzzy sphere and condense, dynamically
locking the horizon gauge field to the boundary value of the external
Maxwell field. The parton current thereby becomes a physical boundary
source for the exterior field.  Our construction replaces the
fictitious membrane of the classical paradigm with a dynamical quantum
membrane populated by microscopic LLL degrees of freedom. The
resulting boundary condition generalizes the classical ingoing
membrane condition and yields explicit quantum, frequency-dependent,
and helicity-dependent corrections, offering a possible probe of the
microscopic quantum structure of the horizon.

\end{abstract}

\end{center}

\newpage
\setcounter{footnote}{0}
%\setcounter{page}{1}
%%%%%%%%%%%%%%%%%%%%%%%%%%%%%%%

\tableofcontents
%%%%%%%%%%%%%%%%%%%%%%%%%%%%%%%

\section{Introduction}

Recently, a microscopic model of quantum black hole was proposed
\cite{Chu:2024qil}.  Like the BFSS \cite{Banks:1996vh}
or the IKKT model \cite{Ishibashi:1996xs}, this
model is based on the quantum mechanics of large $N$ matrices. However,
it differs from them in several essential aspects:
it is not supersymmetric, it contains  
a negative mass term and its fermions transform in the
fundamental representation.
As shown in previous works
\cite{Chu:2024qil,Chu:2024edh,Chu:2026wyh}, and as will be further
supported in the present paper,
these features are not incidental.
Rather, they are necessary ingredients for producing the characteristic
quantum properties of black holes.

The  model has 3 bosonic matrix coordinates
$X^a_{mn}$, $a=1,2, 3$  and $N$ flavor of 2-components
fermion $\psi_{mn\a}, \a =1,2$ in the fundamental representation of
$SU(N)$, with dynamics
governed by the $SU(N)$ quantum mechanics \cite{Chu:2024qil} 
\be \label{L}
S = \int \dd t \Tr \left[
   \frac{1}{2a_0^2 M_P} \dot{X}_a^{2}
   + \frac{M_P}{N^2} \left([X_a,X_b]^2 + 4  X_a^{2}\right)
   %c2 typo before
   + i  \psi^\dag \dot{\psi}
   - a_2 \frac{M_P}{N^2}\psi^\dag \s_a X_a  \psi \right]
%c3 - a_3 r_X M_P .
\ee
plus a possible topological term which we skipped writing here.
Here the $N\times N$  Hermitian matrices $X^a$  represent the
Cartesian coordinates of the 3-dimensional quantized space.
The model has the unitary symmetry
\be \label{UX}
X^a \to U X^a U^{-1}, \quad \psi \to U \psi .
\ee
The model is characterized by three dimensionless coefficients
  $a_0, a_2$ that parameterize the normalization of the bosonic
kinetic term and the Yukawa coupling. 

According to the model, the quantum mechanics of the fuzzy sphere
system reproduces the required macroscopic global charges of the
Schwarzschild black hole \cite{Chu:2024qil} and the rotating Kerr
black hole \cite{Chu:2024edh}, as well as the Bekenstein-Hawking
entropy of black hole through microstates counting.  As a result, we
proposed that the horizon of a quantum black hole is given by a fuzzy
sphere together with a half-filled Fermi sea.  More recently, it was
shown that the model also accounts for the quantum
%c11 Hawking radiation
decay of black hole in terms of the tunneling of fuzzy
spheres \cite{Chu:2026wyh}.  In particular, the semi-classical
decay rate of a black hole \cite{Page:1976df} is reproduced
provided the condition
\be \label{a0}
a_0 = \frac{\pi}{3 }
\ee
is satisfied. This provides a nontrivial check of the model, since
the same
condition has previously been obtained from  matching  the
angular momentum of the quantum rotating fuzzy sphere configuration
with that of the Kerr black hole. Moreover the Hawking temperature
emerges from the quantum mechanical system when an effective
probabilistic description of the tunneling process is adopted.  One of
the motivations of the present work is therefore to ask what other
genuine quantum properties of black holes, and in particular of their
horizons, are predicted by the model.

It is worth mentioning that in the tunneling description of the
%c11 Hawking radiation,
decay of quantum black hole, the monopole plays several essential roles:
(1) It admits fermionic zero modes, which uniquely
select a viable tunneling path from
the otherwise infinite space of possible
bounce configurations. (2) the monopole bounce generates a
tunneling potential with a non-analytic $\log N$ dependence, rather
than the polynomial $1/N$ behavior one might naively expect. This
non-analytic dependence is precisely what is required to reproduce the
black hole decay rate.
Although the monopole configuration was constructed explicitly in
\cite{Chu:2026wyh} as the difference between the initial fuzzy sphere
and the tunneled configuration, its physical origin remained
unclear. In particular, since monopole charge is normally a conserved
quantity, the monopole created during the tunneling
process should arise from a pre-existing monopole already
present in the fuzzy sphere. What is this parent monopole, and where
is it located? Another motivation of the present work is to
answer this question.

A further motivation concerns the nature of the fermionic
degrees of freedom.
It was shown that, when quantized on the fuzzy sphere background, the
fermionic coordinates $\psi$ 
\footnote{
We correct here a typo in the sign of the
kinetic term of the fermion in
\cite{Chu:2024qil}.
} becomes fermionic oscillators with degenerate energy levels.
The corresponding
Fock space forms a Fermi sea, and the half-filled sector gives the
microstate counting responsible for the Bekenstein--Hawking entropy.
These fermionic states may be naturally identified with the partons
anticipated by Susskind \cite{Susskind:1993ws}. Since we identify the
quantum horizon with the fuzzy sphere, we shall refer to them as horizon
partons. This raises a basic dynamical question: can these horizon
partons move, and if so, what transport properties do they generate on
the quantum horizon?

To answer this question, one must understand more precisely the nature of
the Yukawa coupling of the fermions. In the BFSS matrix model
\cite{Banks:1996vh,Taylor:2001vb}, a commutator coupling of the form
$([X_a,\psi]$ gives rise, in the continuum limit, to a covariant
derivative $D_a\psi$ acting on an adjoint matrix field $\psi$
\cite{Taylor:1996ik}. In contrast, the Yukawa coupling in the present
model is not of commutator type. The coordinate matrix $X_a$ acts only
on the left index of $\psi$, while the right index remains a flavor
index. This difference is essential: the fermions do not become ordinary
adjoint fields propagating on the emergent sphere. It is therefore
important to determine what kind of novel horizon dynamics follows from
this non-standard coupling.
In this paper, we find that the nontrivial
topology of the fuzzy sphere implies the existence of an
intrinsic Berry monopole on the fuzzy sphere.
The monopole
has charge $J = (N-1)/2$ where $N$ is the rank of the fuzzy sphere.
When the
fuzzy sphere tunnels to a small fuzzy sphere, conservation of monopole
charge requires the excess monopole charge to be released. This provides
a physical origin for the monopole that appeared in the tunneling
description of
%c11 Hawking radiation
quantum black hole decay process in \cite{Chu:2026wyh}.

The Berry monopole identified here is absent from conventional BFSS
fuzzy-sphere constructions because the BFSS variables are adjoint
matrices. For adjoint degrees of freedom, the Berry connection cancels
between the fundamental and anti-fundamental indices,
and the monopole therefore decouples. This is not the case for the
fermion $\psi$ in the present model,
which transforms as a fundamental fuzzy-sphere degree of freedom.
We show that the fermionic parton states can be identified with the
Lowest Landau Levels (LLL) associated with the Berry monopole.
As a result, their dynamics is
governed by guiding-center motion in the presence of external forces.
It turns out that these partons (which we called $u$ and $d$ in analogy
with Dirac notation) are charged under the electromagnetic field,
so an applied electric field induces Hall drift on the horizon. 
Consequently,
a net Hall current is generated  whenever the black hole
carries an unequal population of (u) and (d) partons, i.e. when the
black hole  is charged.
Moreover, by including  dissipation effect associated with the horizon
temperature, we show that the partons
give rise to an Ohm's current with a
conductivity that is independent of $N$.
We also find a novel polarization-current contribution.
The emergence of these transport properties from the LLL dynamics
of the horizon partons is another main result of this paper.

The transport properties of black holes constitute a fascinating subject. 
According to the seminal insights of the membrane paradigm
\cite{Damour:1978cg,MacDonald:1982zz,Price:1986yy,Thorne:1986iy},
the interaction of the black hole with
external electromagnetic and gravitational fields can be encoded in terms
of
boundary conditions imposed on a timelike stretched horizon.
In particular
the infalling boundary condition for EM field can be
reformulated as the response of a membrane current obeying the Ohm's
law,  with surface conductivity $\s_0 =1/4\pi$.
It is extremely thought-provoking that an apparently empty region
of spacetime can, by virtue of a horizon, obey the Ohm’s law.
Although, in the conventional membrane paradigm,
this current is fictitious and introduced only as a convenient way of
rewriting the infalling boundary condition, its physical interpretation
is highly suggestive and calls for a microscopic identification.
Providing such an identification within our quantum mechanical model
of the horizon is one of the main motivations of this work. In
contrast to the conventional membrane paradigm, the horizon in
our model carries genuine microscopic degrees of freedom. We
find that these degrees of freedom give rise to a real dynamical
electric current containing Ohmic, Hall, and polarization components.
This suggests a quantum membrane paradigm in which the membrane is no
longer merely fictitious, but is instead a physical quantum
system populated by dynamical LLL partons living on the horizon.

Although the derivation of an electrical current carried by the
horizon degrees of freedom is an interesting result,
one essential ingredient remains missing before this can be promoted
to a complete derivation of the quantum membrane paradigm.
The physical system under consideration is a black hole interacting
with an external environment carrying an electromagnetic field,
denoted schematically by $a_A$. 
In matrix model
language, this is modeled by having a big block of matrix, with a
fuzzy sphere subblock $X_a^H$ representing the horizon,
an environmental subblock $X_a^E$, and an off-diagonal
link variables $W_a$, which mediates the interaction between the
two sectors. The parton current derived above is, by itself,
sensitive only to the worldvolume gauge field
$b_A$ living on the fuzzy sphere:
$ j_A = j_A[b_B]$.
However, in order to derive a boundary condition for the external
electromagnetic field from the membrane current, this current must also
be expressible in terms of the external gauge field.
For example the membrane Ohm's law states
$j_A = \frac{1}{4\pi}E_A [a_B]$.
In this paper, we show that the mass term for the
off-diagonal link variables actually becomes tachyonic and 
condenses in a thin region above the fuzzy sphere. The resulting
condensation of the link fields locks the worldvolume gauge field
$b_A$ to the boundary value  $a_A$ of the external gauge field.
This locking mechanism allows the microscopic horizon current to be
translated directly into a boundary condition for the external
electromagnetic field. It thereby promotes the conventional membrane
paradigm into a quantum membrane paradigm: the membrane current is no
longer a fictitious bookkeeping device, but a tangible physical
current carried by microscopic horizon degrees of freedom and
accessible to an outside observer. The microscopic derivation of this
quantum membrane paradigm is the primary result of this work.

A brief summary and plan of the paper is as follows:
In Section 2, we outline the
classical electromagnetic membrane paradigm using the action formulation
developed in \cite{Parikh:1997ma}. To prepare for the description of
a black hole interacting with an ambient Maxwell field, we
consider in Section 3 the
matrix model expanded around
two types of backgrounds:  a fuzzy sphere background (a black hole block)
and a commuting flat space background (an
environmental block). We then develop the
large $N$ continuum descriptions
of the corresponding matrix model dynamics.
The dynamics of the fermions on the
fuzzy sphere horizon is more novel and is analyzed in detail 
in Section 4. We show that the  parton states are in fact
Lowest Landau Level (LLL) states associated with
an intrinsic Berry monopole
on the fuzzy sphere, and  derive their electric response. In
particular, we show both a Hall current, arising from the LLL
guiding-center dynamics, and an Ohmic current, arising from dissipative
effects of the thermal horizon. In Section 5, we systematically analyze
the interaction between background blocks in the matrix model in terms
of off-diagonal link variables.
In Section 6, we derive a Higgs-like
locking mechanism of the horizon gauge field
with the environmental Maxwell gauge field. We show that
the link field develops a tachyonic
instability in a thin region outside the fuzzy sphere and condenses.
This condensate locks the two gauge fields, thereby
converting the electrical response of the horizon
partons into a physical boundary condition for the external Maxwell
field. Our construction therefore gives a concrete microscopic origin to
the membrane paradigm, replacing its fictitious surface current by the
transport response of physical LLL degrees of freedom living on the
horizon. 
Moreover, we show that the novel membrane currents give the fuzzy
horizon a quantum reflectivity. This opens a possible observational
window onto the microscopic structure of the horizon and
provides a potential test of quantum gravity.

\section{Classical Electromagnetic Membrane}

The membrane paradigm is a remarkable
framework for black hole physics that
replaces the actual event horizon with a fictitious stretched horizon
— a timelike surface located just outside the true horizon — endowed
with effective electromagnetic/fluid properties that reproduce,
for external
observers, the same physics as the standard
formulation of general relativity.
With deep insights,
the membrane paradigm \cite{Damour:1978cg,MacDonald:1982zz,
  Price:1986yy,Thorne:1986iy} can be derived from
an action formulation \cite{Parikh:1997ma}
which is particularly transparent.
Consider a Maxwell theory
\be \label{S-EM-bulk}
S_{\rm out} = \int d^4 x \sqrt{-g}
\left(-\frac{1}{16\pi} F_{\m\n}^2 + J^\m A\mu \right)
\ee
in the spacetime outside the black hole. Instead of considering the null
surface
of the horizon, the membrane paradigm replaces the actual event horizon
with a timelike fictitious stretched horizon sitting right outside the
black hole. The authors of \cite{Parikh:1997ma}
discuss that with the inclusion of a
surface term
\be
S_{\rm surf} = \int d^3 x \sqrt{-h} \; j^A A_A, 
\ee
the boundary term of \eq{S-EM-bulk} can be canceled, giving a surface
current from the boundary value of the bulk gauge field
\be \label{js}
j_A = \frac{1}{4\pi} F_{A \m}n^\m,
\ee
where $n$ is the normal to the surface. Here the indices $a=1,2,3$
denotes the Cartesian components. The indices $A$ are the tangential
indices along the spherical horizon.

Central to the membrane paradigm is the adoption of a stretched horizon
located just outside the horizon of the  Schwarzschild metric
\be \label{sch-metric}
ds^2 = -f(r) dt^2 + \frac{dr^2}{f  (r)} +r^2 d\Omega^2, \qquad
f(r) = 1- \frac{R}{r}, \quad R=2GM
\ee
and the assumption that the membrane
current for a fiducial observer (FIDO) living on the stretched horizon
satisfies the Ohm's law
\be \label{jE}
j_A = \frac{1}{4\pi} E_A.
\ee
Here the electric field $E_A = F_{A \tau}$ 
is defined with respect to the FIDO proper time $d \tau = \sqrt{f} dt$
giving
\be \label{EFf}
E_A = F_{At} \frac{1}{\sqrt{f}}.
\ee
The important point of \eq{jE} is that
Ohmic surface conductivity
is universal and independent of black hole mass (and is equal to $1/4
\pi$ in the Gauss unit).
The surface current \eq{jE} is fictitious in the
membrane paradigm
and is simply an elegant reformulation of the ingoing 
boundary condition  at the horizon.
In fact, eliminating $j$ from \eq{js} and \eq{jE} gives
\be\label{EAFA}
F_A{}^n = E_A.
\ee
Using $n^\m= \sqrt{f} \d^\m_r$ 
for the outward unit normal of the stretched horizon, we have
\be \label{Fnr}
F_A{}^n = \sqrt{f} F_A{}^r.
\ee
For  fields without angular dependence ($S$-wave), 
we obtain from \eq{EAFA}
\be \label{fAtA}
f  \del_r A_A = \del_t A_A
\ee
in the gauge $A_t=0$.
This is nothing but simply
\be \label {ingoingBC}
(\del_t - \del_{r_*} )A_a = 0,
\ee
where $dr_* = dr/(1-R/r)$ is the tortoise coordinate.  
This means
\be
A_a = A_a (t+r_*)
\ee
is entirely ingoing at the stretched horizon.
In the frequency domain, it is
\be
A_a = A_a (\o) e^{-i \o (t+ r_*)}
\ee
and there is no outgoing wave component $e^{-i \o (t - r_*)}$.

%% In terms of the tortoise coordinate
%% the Schwarzschild metric reads
%% \be
%% ds^2 = -f(r) (dt^2 - dr_*^2)  +r^2 d\Omega^2, \qquad
%% f(r) = 1- \frac{2GM}{r}, \quad dr_* = \frac{dr}{f} ,
%% \ee
%% and for a stretched horizon surface located at $r = R + \d$,
%% we have $n^\m=  \d^\m_{r_*}/ \sqrt{f}$ 
%% for the outward unit normal and
%% \be \label{js1}
%% j_A = \frac{1}{4\pi} F_{A r_*} \frac{1}{\sqrt{f}}.
%% \ee
%% Central to the membrane paradigm is the assumption that the membrane
%% current satisfies the Ohm's law
%% \be \label{jE}
%% j_A = \frac{1}{4\pi} E^{\rm local}_A.
%% \ee
%% This means the black hole horizon
%% is a conductor with a Ohmic surface conductivity of $1/4 \pi$.
%% Note, however, that the surface current \eq{js1} is ficitious in the
%% membrane paradigm
%% and is simply an elegant reformulation of the ingoing 
%% boundary condition  at the horizon.
%% To see this, we note that
%% the local electric field $E_A = F_{A \tau}$
%% is defined with respect to the local time $\tau$, which gives
%% \be \label{jE1}
%% j_A = \frac{1}{4\pi}F_{A t}\frac{1}{\sqrt{f}}.
%% \ee
%% when written in terms of the Schwartschild time $t$.
%% Equating \eq{js1} and \eq{jE1}, the factor of $\sqrt{f}$ cancels and
%% for fields without angular dependence ($S$-wave), we obtain
%% \be \label {ingoingBC}
%% (\del_t - \del_{r_*} )A_a = 0
%% \ee
%% in the gauge $A_t=0$.
%% This means
%% \be
%% A_a = A_a (t+r_*)
%% \ee
%% is entirely ingoing at the stretched horizon.
%% In the frequency domain, it is
%% \be
%% A_a = A_a (\o) e^{-i \o (t+ r_*)}
%% \ee
%% and there is no outgoing wave component $e^{-i \o (t - r_*)}$.

We emphasize that in the classical membrane paradigm, there are
no extra degrees of freedom or charges put on the black hole, the
surface current is simply introduced to allow for a reformulation of
the ingoing boundary condition of black hole as a transport property
of the horizon.  The current is however fictitious and carries no
independent physical information.  The situation is drastically
different in our quantum mechanical model of black hole.
According to model, a quantum black hole is made up of a fuzzy
sphere together with a sea of fermi degrees of freedom residing on the
horizon.  In this paper, we show that the set of fermi states not only
provides the microstates counting of the Bekenstein-Hawking entropy,
but they are also charged and give rise to an electric current on the
fuzzy sphere horizon. We derive that apart from an Ohmic component,
the current also receives a Hall component.

We remark that the Ohm's law may be seen as 
a holographic relation for a black hole
where the radial properties of incoming modes
can be encoded in terms of some transport properties of the boundary.
In the original membrane paradigm, this bulk-boundary relation is
anticipated with the introduction of a non-dynamical membrane current.
In section 6 of this paper,
we derive the Ohm's law through a novel mechanism of
locking. Thus, different from
the well known bulk-to-boundary propagator formalism
\cite{Witten:1998qj,Gubser:1998bc},
our model provides another
interesting mechanism of how holographic relation may emerge
from the boundary,

\section{Emergence of EM on  Fuzzy Sphere and Cartesian Environment}

To describe the interaction of the horizon with the bulk Maxwell
field outside the black hole,
it is necessary to consider the covariantized form of the
matrix model \eq{L}, in which ordinary time derivatives are replaced
by covariant derivatives. This is important because Maxwell theory
possesses an exact gauge symmetry, which cannot emerge from a matrix
model containing only a global symmetry at the fundamental level.
In practice, we work in the gauge-fixed formulation by imposing the
temporal gauge $A_0=0$. Nevertheless, the corresponding Gauss-law
constraint must be imposed. We will return to this point in
Section \ref{sub-Gc},
where we also explain why the net parton charge can be
identified with the electric charge of the black hole.

\subsection{Fuzzy sphere block}

In this section, we consider the
bosonic part of the model where the emergence of the gauge system 
from the matrix quantum mechanics
is more or less standard. However, the treatment of the
fermionic part of the model is more novel and it will be considered in the
next section. Consider a fluctuation
\be \label{XqB}
X_a=J_a+ \d X_a, \quad \d X_a = R B_a
\ee
where $J_a$ is the fuzzy sphere background 
\be \label{FSB}
[J_a,J_b]=i \epsilon_{abc}J_c, \qquad J_a^2 = J(J+1) 
\ee
for spin $J$
\be
J = \frac{N-1}{2}.
\ee
The action for the fluctuation is given by  (${\rm H}$ stands for horizon)
\be \label{SBB}
S_\HH =   R^2 \int dt \; \Tr_N \left[
  \frac{1}{2 a_0^2 M_P} \dot{B}_a^2 -\frac{2 M_P}{N^2}B_a \cN_{ab} B_b
\right],
\ee
where
\be
\cN :=  (S\cdot L)^2 - (S\cdot L)  -2, \qquad  (S_c)_{ab} := -i \e_{cab}, 
\ee
is the quadratic operator of fluctuations. To evaluate the action \eq{SBB},
we
note that $\Tr B_a \cN_{ab} B_b = \Tr (TB)^2 -\Tr BTB -2\Tr B^2$ where
$T:=S\cdot L$, so we only need to compute $TB$.
Now in the large $N$ limit,
one may decompose $B_a$ into components $b_A$
tangential to the sphere and a scalar part
$\upvarphi_{\rm H}$
using the Killing tangent frame (see appendix)
\be \label{Bbphi}
B_a = K^A_a b_A + n_a \upvarphi_{\rm H}.
\ee
At the same time, the angular momentum becomes 
$L_c = -i K_c{}^A D_A$ where $D_A$ is the covariant derivatives on the unit
sphere. It is easy to show that
\be
(TB)_a = K_a{}^A (\ve_A{}^BD_B \upvarphi_{\rm H} -b_A) +n_a
(f-2\upvarphi_{\rm H}),
\ee
where
\be
f_{AB} := D_A b_B- D_B b_A , \qquad f: = \frac{1}{2}\ve^{AB} f_{AB}.
\ee
Using
\be \label{TrN}
\Tr_N = \frac{N}{4\pi} \int d\O
\ee
in the large $N$ limit, we obtain
\be
\Tr_N B_a \cN_{ab} B_b = \frac{NR^2}{4\pi}\int d\O \left(F^2 +
(\nabla_A \upvarphi_{\rm H})^2 + \frac{4}{R^2} \upvarphi_{\rm H}^2
- \frac{8}{R} F \upvarphi_{\rm H} \right),
\ee
where $\nabla_A = \frac{1}{R} D_A$ is the covariant derivative on $S^2_R$
and $F_{AB} = \frac{1}{R} f_{AB}$, $F= \frac{1}{2}\ve^{AB} F_{AB}$.  
To obtain the standard Maxwell form of action with equal coefficients for
the electric term and the magnetic term, we introduce a rescaled time
\be \label{cN}
\tilde{t} = c_N t, \qquad c_N = \frac{2 a_0 M_P R}{N}.
\ee
Note that $c_N$ is independent of $N$.
As a result, we obtain the bosonic action for the fuzzy horizon
\be \label{SH1}
S_\HH = \frac{1}{2 \tilde{g}^2}\int d\tt d\S
\left[
  (\del_\tt b_A)^2 -F^2 + (\del_\tt  \upvarphi_{\rm H})^2 -
  (\nabla_A  \upvarphi_{\rm H})^2 -  \frac{4}{R^2} \upvarphi_{\rm H}^2
+ \frac{8}{R} F \upvarphi_{\rm H}
\right],
\ee
where
\be
\frac{1}{2 \tilde{g}^2} := \frac{R}{4 \pi a_0}
\ee
is the gauge coupling for the fuzzy sphere system and $d\S =  R^2 d\O$
is the volume element of $S^2_R$. Note that there is no mass term for
the vector
fluctuation $b_A$ even through the original action \eq{L}
has a (negative) mass term. This
is due to the special property of the quadratic fluctuation operator
$\cN_{ab}$.

We note that the action \eq{SH1} describes
the dynamics of the vector fluctuations
over the fuzzy sphere in terms of the original time $t$ (up to a
proportional constant $c_N$) of the matrix model.
As $t$ was the time used to measure the
energy of the fuzzy sphere system which was matched with the mass of the
black hole, $t$ and $\tt$ should be identified with
the asymptotically flat Schwarzschild time coordinate.
As discussed above, the Ohm's law is formulated for the FIDO and is
written in terms of
the FIDO proper time $\tau$. So let us also write down the gauge theory
on the sphere using the FIDO time. To identify the desired time scale.
Let us note that $\tau$ is defined by
\be
d \tau = \sqrt{f_{\rm sh}} dt, \qquad 
f_{\rm sh} := f|_{r=R+\d},
\ee
where $f_{\rm sh}$ is the metric factor of the Schwarzschild metric
at the location of the stretched horizon $r= R +\d$.
Now the proper distance $\ell$ of the stretched horizon from the horizon
is given by
\be
\ell = \int_R^{R+\d} \frac{dr}{\sqrt{1-R/r}} \approx \sqrt{2\d R}
\ee
for small $\d/R$. Therefore for a stretched horizon located at
$\ell \sim l_P$,
we have  $\d \sim \ell^2/R$ and $f_{\rm sh}\sim 1/N^2$. Therefore
the FIDO time is proportional to the Schwarzschild time by a factor of
$1/N$:
\be \label{FIDO-time}
\tau \sim \frac{t}{N}.
\ee
The relation holds up to an order 1
coefficient that depends on the location of the
stretched horizon.
For simpler looking formulae, let us consider a change of time
with  specific coefficient as below:
\be \label{FIDO-time2}
\tau := \frac{2a_0 M_P l_P}{N} t.
\ee
A change of the numerical coefficient in \eq{FIDO-time2}
modifies only the relative coefficients of terms in \eq{SH2} below in an
$N$-independent manner and does not affect our analysis or results. 
With \eq{FIDO-time2}, we obtain
\be \label{SH2}
S_\HH = \frac{R^2}{2 g_{\rm BH}^2}\int d\t d\O
\left[
  (\del_\tau b_A)^2 + (\del_\tau  \upvarphi_{\rm H})^2 - \frac{1}{l_P^2}
  \left(f^2+ (D_A  \upvarphi_{\rm H})^2 +4 \upvarphi_{\rm H}^2
- 8 f \upvarphi_{\rm H}\right)
\right],
\ee
where 
\be
\frac{1}{2 g_{\rm BH}^2} = \frac{l_P}{4\pi a_0}.
\ee
Note that the action \eq{SH2} written in the FIDO frame is very simple:
apart from a simple area geometric area factor
in front, it is entirely constructed from the metric $h_{AB}$ and covariant
derivative $D_A$ of the unit sphere, and so the
dynamics of $b_A$ and $\upvarphi_{\rm H}$ are explicitly $N$ independent.

\subsection{Cartesian environmental block}

To describe the exterior spacetime outside the black hole in matrix model,
we consider
the theory of \eq{L} with a rank $N'$ and consider an ansatz
\be \label{XLY}
X_a = L Y_a.
\ee
Here $L$
is a regulator length that characterizes the size of the environmental
space outside the black hole.
Since we are interested in the interaction between the black hole block
and the environmental block, therefore let us use the same FIDO time
\eq{FIDO-time2} and derive the continuum description.
Substituting \eq{XLY},
the action for the environmental block reads
\be\label{S-env0}
S_{\rm env} = \int \dd \tau
 \frac{l_P L^2}{a_0 N } \Tr_{N'} \left[ 
  (\del_\t Y_a)^2 + \frac{L^2N^2}{2 N^{'2}l_P^2}
  ([Y_a,Y_b]^2  + \frac{4}{L^2}Y_a^2 )
\right] + S_{\rm F},
 \ee
 where $S_{\rm F}$ denotes the fermion action, which we will show later
 that in the large $L$ limit,  $\psi$ decoupled and can be ignored.
As $N'$ characterizes the number of degrees of freedom of the
 environmental
block, we will take both $N'$ and $L$ to
be  infinity. 
Note that in the $L \to \infty$ limit,
the mass term $\sim 1/L^2$ is dropped. The theory has the
usual Yang-Mills type matrix equation of motion and
admits the bosonic background, 
\be
 [y_a, y_b] =0.
 \ee
To  get a gauge theory, let us consider
\be
Y_a = y_a +  \frac{R}{L}A_a
\ee
with a perturbation $A_a$. The coefficient in front of $A_a$
is chosen such that the perturbation matches that of \eq{XqB} which 
allows to display the locking relation \eq{locking-ba}
of gauge fields in the most transparent manner
without any relative coefficients. As for $y_a$,
it is customary to take
it to be a momentum variable $p_a := - i \del/\del x_a$. 
 For reasons which will be clear below, we
will take it to be a shifted momentum: 
\be
y_a = \frac{x_a}{2 l_P L} - i \frac{\del}{\del x_a},
\ee
where $x_a$ plays the role of the coordinates for
the emerged 3 dimensional space. The coefficient is chosen so that,
for the vacuum configuration, the physical coordinate is
$x_a = X_a 2l_P$, obeying the same uniform normalization convention as
for the fuzzy sphere. 
For fluctuation $A_a = A_a(x)$, we obtain   in the large $N'$ limit,
\be
   [Y_a,Y_b] = -\frac{iR}{L} F_{ab},  \qquad
   F_{ab} = \del_a A_b - \del_b A_a.
   \ee
   Also, in order to maintain the standard
   electric and magnetic coefficients for a gauge theory, we take
   \be \label{LNPl}
L = \frac{N'l_P}{N}.
   \ee
In the the large $N'$ limit, the trace is converted into an integral
   \be \label{rhoV3}
\Tr_{N'}f = \frac{N'}{V_3}\int d^3 x f(x),  
\ee
where $V_3$ is a regularized volume of space.
The normalization factor is fixed by requiring
equality for the case of $f=1$. To
make the gauge coupling finite, we take
\be \label{V3bL}
V_3 = \nu_3 R^2 L
\ee
with $\nu_3$ a $N$-independent constant.
As a result, we obtain
\be \label{S-env}
S_{\rm env} =\frac{1}{2g_{\rm EM}^2}
\int\dd \tau\,\dd^3x
\left[
 (\partial_\tau A_a)^2 - \frac{1}{2}F_{ab}^2 \right],
\ee
where 
\be
\frac{1}{2 g_{\rm EM}^2} = \frac{1}{a_0 \nu_3} 
\ee
is finite and independent of $N'$. In particular, 
the Gauss-unit normalization of \eq{S-EM-bulk} is obtained if $\nu_3 =24$.

Note that the unitary symmetry \eq{UX} of the matrix model
implies the gauge symmetry
\be
A_a \to A_a - \del_a \lambda
\ee
for $U = e^{i \l}$. 
For later
consideration, we can also decompose the gauge field at the horizon $r=R$
into tangential components $a_A$ and a  scalar part
$\upvarphi_{\rm E} $ (${\rm E}$ stands for environment)
using the Killing tangent frame
\be
A_a = K^A_a a_A + n_a \upvarphi_{\rm E}.
\ee
The corresponding gauge transformation is
\be
\d a_A = D_A \l, \qquad \d \upvarphi_{\rm E} =0.
\ee
Finally,
we note that normally it is natural to take the regularized volume to
scale as $V_3 \sim L^3$ if the 3-space under
consideration is homogeneous.
On  the other hand,  a thin layer of space of a proper
thickness $\ell \ll R $ above the horizon of the Schwarzschild metric
has the proper volume
\be \label{V-proper}
V = 4\pi R^2 \ell.
\ee
Our choice of $V_3 \sim R^2 L$ respects this distinction of
the radial and the angular directions and 
is compatible with
the physical picture that the space outside the horizon is dictated by
the Schwarzschild metric \eq{sch-metric}.

Finally, we note that $S_{\rm F}$ in \eq{S-env0}  is given by
\be
S_{\rm F} = \Tr_{N'} \left[
  i \psi^\dag \del_\t \psi - \frac{a_2 }{2 a_0 M_P l_P N'}
\psi^\dag \s_a Y_a \psi \right]
\ee
where we have used \eq{LNPl}.
In the  large $N'$ limit with \eq{rhoV3}, one can rescale $\psi$ to keep
the kinetic term finite,  however  the Yukawa term
drops out. Therefore $\psi$
decoupled in the limit and we have ignored it in \eq{S-env}.

\section{Dynamics of  Partons on Fuzzy Sphere}

\subsection{Charged black hole from partons ensemble}

It was shown in \cite{Chu:2024qil} that the fermions
\be \label{LF}
L_F = \Tr \left[
  i \psi^\dag \dot{\psi} -
  \frac{a_2 M_P}{N^2}\psi^\dag \s_a X_a \psi \right],
\ee
when
quantized on the fuzzy sphere background \eq{FSB}
gives rise to a set of $2N^2$
fermionic oscillators
\footnote{
We have renamed $\xi, \chi$ of \cite{Chu:2024qil} by $u,d$ here. 
}
$u^p_k$ and $d^p_k$, $p, k = 1, \cdots, N$:
  \be \label{xi-chi}
  u^{p}_k := \cU^{p \dag}_{n\b} \psi_{nk\b}, \quad
  d^{p\dag}_k := \cV^{p \dag}_{n\b} \psi_{nk\b}
  \ee
Schematically, we have decomposed 
the two components spinor $\psi_\a$ into the form
\be
\psi = \begin{pmatrix}
  u \\
  d^\dagger
  \end{pmatrix}
\ee
which resembles the Dirac fermion in 4 dimensions. In \cite{Chu:2024qil},
we have
noted that the $u, d$ oscillators play precisely the role of partons
which has been conjectured \cite{Susskind:1994vu} as a
mean to account for the area law of
the Bekenstein-Hawking entropy. In our model, the partons have become
dynamical degrees of freedom obeying a free (normal ordered)
Hamiltonian 
\be\label{HF-diag}
H_F = \frac{a_2 M_P}{2N} (r+s),
\ee
where
\be \label{rs}
r :=  \sum_{p,k=1}^N    u_k^{p\dag} u_k^{p}, \quad
s:= \sum_{p,k=1}^N d_k^{p \dag} d_k^{p} 
\ee
are the occupation number for the $u$ and $d$ states.
Strictly speaking, coupling the orbital spin $J$
to the local spin-1/2 basis splits the two radial-spin
sectors into bundles of degeneracy $N+1$ and $N-1$ per flavor.
In the following we keep
only the leading large-$N$ degeneracy $N$ per flavor;
the exact shift affects only subleading $1/N$
corrections to the entropy and transport coefficients.
It was proposed to identify a quantum black hole with a fuzzy sphere
endorsed with an ensemble of states with
\be \label{r_p_s}
r+s =N^2, 
\ee
i.e. a half-filled Fermi sea.

Now, as the black
hole quantum mechanics \eq{L} has a $U(1)$ symmetry
\be
\psi \to e^{i \a} \psi.
\ee
This implies that $r-s$ is conserved and means that
we can further refine the ensemble of states to one with constant
$r-s$. The analogy with the Dirac fermion suggests to identify
\be \label{Qrs}
Q := q (r-s)
\ee
as the charge of a configuration of fermions where  $\pm q$ is the charge
of the individual $u$ and $d$ partons. $q$ specifies the coupling
of the partons to the EM field and will be fixed below.
We propose to identify a fuzzy sphere with an ensemble
of fermi states characterized by
\be \label{rsNQ}
r= \frac{1}{2}(N^2 + n), \quad s = \frac{1}{2}(N^2 - n)
\ee
with a black hole of size
$R= N l_P$, charge $Q = nq$ and an energy
\be \label{EQ}
E = \frac{R}{2G} + \frac{Q^2}{2R}.
\ee
Note that here $R$ denotes the outer horizon radius.
As the inner horizon radius must be less than $R$,
we have $|Q| \leq R$. 
This implies that $Q$ is at most of order $N$.

To justify this claim, let us first examine the entropy counting.
(i) For an
ensemble of fermi states with $r,s$ given by \eq{rsNQ}, there is a
degeneracy of
\be
\Omega = \binom{N^2}{r}^2.
\ee
We obtain for the microstates entropy $S = \log \Omega$ in the leading
order of $N$ large
\be \label{SQ}
S= 2N^2 \left( 1 + O (\frac{n^2}{N^4}) \right).
\ee
Therefore as long as $Q$ is small enough with $n \ll N^2$,
\eq{SQ}
reproduces the Bekenstein-Hawking entropy in the leading order
of large $N$.
We remark that for the neutral Schwarzschild black hole considered
in \cite{Chu:2024qil},
the counting of microstates was performed without imposing
the condition $r-s =0$. The Bekenstein-Hawking entropy was also
reproduced in the leading order of large $N$. From now on, we will
identify a quantum black hole with a half filled
\eq{r_p_s} Fermi sea together with the conserved $U(1)$ charge condition
\eq{Qrs} imposed.
(ii) Next, for the energy. For the ensemble of states 
\eq{r_p_s} of the half-filled Fermi sea,
they contribute an energy \eq{HF-diag} as before, and hence
first term of \eq{EQ} is reproduced
from the sum of the bosonic and fermionic energy of
the quantum mechanics. For the second term, it
is simply the electrostatic energy
of  a thin shell of charge $Q$ of radius $R$. Therefore it can be
obtained from
our model provided that the partons act correctly
as the source of a bulk Maxwell field.  This last point is important and
we will need to show
that the partons couple on the fuzzy sphere source
the external Maxwell field correctly. This will be done in two different
ways:
We will spell out in this section
how the partons coupled to the worldvolume gauge
field. Then we will show in section 6 that the worldvolume gauge field
is locked
to the gauge field in the bulk spacetime near the horizon. In this way, the
horizon partons do source the bulk Maxwell theory and gives rise to the
electric energy term in \eq{EQ}.
The same conclusion can also be arrived with an analysis of the
Gauss laws.  In section 5.3 we show that the parton charges source
the Gauss law of the external Maxwell field. That the same result
is obtained is a demonstration of the internal
consistency of our model.

\subsection{Fundamental fuzzy sphere state and Berry monopole }

We recall that in \cite{Chu:2026wyh},
%c11 Hawking radiation
the  decay of a quantum  black hole was described as a
tunneling process of the fuzzy-sphere background accompanied by a
monopole creation.
Although the monopole configuration was constructed explicitly,
it was not clearly demonstrated how such a monopole can be created
since monopole charge is normally a conserved 
quantity in a gauge theory. A related question concerns the role of
the fermionic partons. If the black-hole horizon is identified with
the fuzzy sphere, then, beyond contributing to the energy and
microstates counting of the quantum black hole, how do these partonic
degrees of freedom participate in the dynamics of the quantum horizon?

In the following, we show that the partons on the fuzzy sphere are not
ordinary propagating two-dimensional fermions. Rather, they are
Lowest-Landau-Level (LLL) degrees of freedom associated with a Berry
monopole (see appendix C).
Since their kinetic energy is quenched, the
partons do not exhibit conventional second-order propagation on the
sphere. Their dynamics is instead first order and is governed by
guiding-center motion in the background monopole field, together with
their coupling to the fuzzy-sphere gauge field. Thus, a localized
parton wave packet may drift on the fuzzy sphere in response to
tangential electric fields or background potentials, but it does not
propagate ballistically like an ordinary two-dimensional fermion. We
further argue that this partonic LLL sector supplies the microscopic
degrees of freedom responsible for making the fuzzy sphere
electrically and dynamically responsive, thereby turning the classical
fiducial membrane into a dynamical quantum membrane.

Let us begin with the following observation.
Although the fermionic variable $\psi$ in our matrix model
is a $N\times N$ matrix as in the BFSS model,
its Yukawa coupling to the bosonic coordinates takes the form
\footnote{
We are not aware that this type of matrix coupling and its continuum
interpretation has
been considered before.
}
\be \label{yukawa}
\Tr \psi^\dag \s_a X_a \psi
\ee
rather than the  usual commutator form
$  \Tr \psi^\dag \s_a [X_a, \psi]$ as in the
BFSS theory. It is well known that 
the BFSS type commutator coupling gives rise, in the large $N$ limit,
to a spatial derivative term on the emergent geometry,
and is therefore appropriate for describing ordinary
propagating degrees of freedom. On the other hand,
the Yukawa coupling \eq{yukawa} in our model was designed
to produce the ground-state degeneracy needed to account for the
Bekenstein--Hawking entropy in the model \cite{Chu:2024qil}.
In particular,
we note that 
the second index $p$ of $\psi_{mp}$ does not participate in the interaction
with $X_a$ and so it is actually
a flavor index with respect to the spatial coupling. As such, we have $N$
flavors of vector fermions $\psi_m^{(p)}$, $p = 1, \cdots, N_f =N$.
In general, a $N_1 \times N_2^*$ representation of $SU(2)$
can be represented as a rectangular $N_1 \times N_2$ matrix and
the action of $SU(2)$  is given by
\be
L_a \psi = J^{(N_1)}_a \psi - \psi J^{(N_2)}_a,
\ee
where $J^{(N_\ri)}_a$ is a $N_\ri\times N_\ri$ representation of $SU(2)$. The
includes  BFSS case for adjoint matrices. For our vector fermions, 
we have (flavor index $p$ hidden)
\be \label{Lpsi}
(L_a \psi)_m = (J_a)_{mn} \psi_n 
\ee
and so \eq{yukawa} is simply the spatial part of the Dirac operator
$D= i \del_t + \frac{1}{R} \s_a J_a$ on the fuzzy sphere for a fundamental
fermion.

In the appendix C, we show that the $SU(2)$ algebra underlying the fuzzy
sphere naturally gives rise to a Berry monopole
\be \label{berry-c1}
\cA = -i \bra{\O} d \ket{\O} = J(1-\cos \th) d\phi,
\ee
The associated Berry
phases cancel for adjoint scalar matrices, so they are neutral under
this monopole, whereas fundamental fuzzy-sphere states carry a nonzero
monopole charge. Consequently, the fermionic states in our model
couple directly to the Berry monopole background.  In fact
there are precisely the LLL of the monopole. To see this, let us
consider a fundamental fuzzy-sphere state transforming in the spin $J$
irreducible representation of $SU(2)$
\be
\ket{\psi} := \sum_{m =-J}^J \psi_m \ket{J,m}.
\ee
We can use the coherent state (see appendix) to turn it
into a field on $S^2$:
\be
\psi(\Omega) := \la \Omega|\psi \ra = \sum_{m =-J}^J  \psi_m\la\Omega|J,m \ra. 
\ee
Now, the fact that there is a phase freedom \eq{gt-Omega}
in the definition of
the coherent state induces the transformations for the field $\psi(\O)$
as
well as the Berry connection:
\be
\psi(\O) \to e^{-i\b(\O)} \psi (\O), \quad \cA \to \cA + d\b.
\ee
This means $\psi(\O)$ is of charge 1 with respect to the Berry
connection. Now it is 
\be \label{zz}
\la\Omega|J,m \ra = \sqrt{\binom{2J}{J+m}}\frac{z^{J+m}}{(1+|z|^2)^J}
\ee
where we have used, for the north patch, the
stereographic coordinate $z = e^{i\phi} \tan \frac{\th}{2}$.
This means $\psi$ is a LLL wavefunction
\be
\psi(z,\zb) = \frac{P_{N-1}(z)}{(1+|z|^2)^J},\qquad
\mbox{where $P_{N-1}(z)$ is a polynomial of degree  $\leq N-1$},
\ee
and satisfies the covariant holomorphic equation
\be
D_{\zb} \psi =0,
\ee
with
\be \label{A-z}
\cA_\zb = iJ\frac{z}{1+|z|^2}, \quad \cA_z = -iJ\frac{\zb}{1+|z|^2},
\ee
which, in polar coordinates,  is precisely the Berry connection
\eq{berry-c1}.

\subsection{LLL partons on fuzzy sphere}

Let us now come back to our model and
write the matrix fermion as $N$ column vectors indexed by a
flavor index $I$,
\be
\psi_{\a;mI}, \qquad m =1, \cdots, N, \quad I = 1, \cdots, N_F =N.
\ee
To facilitate the discussion of the large $N$ continuum limit,
let us write the action in terms of spin states
of the fuzzy sphere.
For each fixed $\a, I$, define
\be
|\psi_{\alpha I}\rangle
=
\sum_{m=-J}^{J}\psi_{\alpha;mI}|J,m\rangle,
\ee
the fermionic action \eq{LF} in the FIDO time can be written as
\be
S_F =\int d\t\; \sum_{I=1}^N
\left[ i \la \psi_{\a I} | \dot{\psi}_{\a I} \ra
  - \k \bra{\psi_{\a I}} \s_a J_a \ket{\psi_{\a I}}
  - q \bra{\psi_{\a I}} \s_a B_a \ket{\psi_{\a I}}
  \right]  \ee
where
$\kappa :=\frac{a_2}{2a_0R}$ and
\be q:=\frac{a_2}{2a_0}.
\ee
Note  that the parton charge is independent of $N$ and is universal
to all black holes.
Now use the spin-$J$ coherent state and define
\be
\chi_{\alpha I}(\Omega,\tau)
=
\sqrt{\frac{N}{4\pi R^2}}\,
\langle\Omega|\psi_{\alpha I}(\tau)\rangle.
\label{continuum-field}
\ee
over the  physical sphere $S_R^2$ and substitute \eq{Bbphi},
we obtain  leading the continuum action
\be \label{SF1}
S_F
=
\int\dd\tau\,\dd\S\,
\Big[\sum_{I=1}^{N} i\chi_I^\dagger\partial_\tau\chi_I
  - j_A b_A - n_P \mu
\Big].
\ee
where
 \be
 j_A:= q K_a{}^A \sum_{I=1}^{N} \chi_I^\dagger\sigma^a \chi_I,\qquad
n_P:=\chi_I^\dagger (\sigma \cdot n) \chi_I \qquad
\mu := q \upvarphi_{\rm H} + \e_0 
\ee
are electric current,
parton number density and  chemical potential respectively, 
and the constant $\e_0$ is given by 
\be
%c9 typo
\e_0 :=\frac{a_2}{4a_0l_P}.
\ee
%c13
This is the energy scale  which control the creation of partons and the
associated virtual process.
We have substituted the decomposition \eq{Bbphi} for $B_a$.
%% We remark that
%% the second term above arises in the large $N$ limit of the term $\s_a J_a$
%% in the matrix quantum mechanics,
%% whose  exact diagonalization has been considered in \cite{chu} by considering
%% an characteristic equation. The same result can also be obtained by
%% considering the total angular momentum
%% \be
%% \bm K =\bm J+\bm S,
%% \qquad
%% S=\frac12.
%% \ee
%% one has
%% \be
%% \sigma\cdot J
%% =K^2-J^2-\frac34.
%% \ee
%% The two exact multiplets are $K=J\pm\tfrac12$.  In the strict
%% large-$J$ approximation, we retain only the leading terms
%% \be
%% \sigma\cdot J\simeq \pm J,
%% \qquad
%% \dim\mathcal H_\pm\simeq 2J,
%% \qquad
%% 2J\simeq N.
%% \label{largeJ-bands}
%% \ee
That $j_A$ represents an electric current can be seen from the fact that
it satisfies a conservation law
$\del_\t \rho + \nabla_a j^a = 0$
on the fuzzy sphere, where
\be
\rho := q \sum_{I=1}^{N}\chi_I^\dag \chi_I
\ee
is the charge density for the partons with the charges of $u$ and
$d$ given by $\pm q$, and
%c9: \cL_a is of dim 1/length
$\nabla_a = i \cL_a$ is
the rotational derivative. 

Next let us introduce the particle-hole parton variables directly.
Introduce position-dependent spin frame
that diagonalizes $\sigma\cdot n$:
\be
(\sigma\cdot n)\eta_\pm=\pm\eta_\pm.
\label{eta-def}
\ee
For the polar angle parameterization,
\be\label{eta_pm}
n_a(\theta,\phi)
=(\sin\theta\cos\phi,\sin\theta\sin\phi,\cos\theta),
\ee
we have 
\be
  \eta_+
=
\begin{pmatrix}
\cos(\theta/2)\\
e^{i\phi}\sin(\theta/2)
\end{pmatrix},
\qquad
\eta_-
=
\begin{pmatrix}
-e^{-i\phi}\sin(\theta/2)\\
\cos(\theta/2)
\end{pmatrix}
\ee
around the north pole. 
They satisfy
\be
\eta_s^\dagger\eta_{s'}=\delta_{ss'},
\qquad
\eta_\pm\eta_\pm^\dagger
=\frac{1\pm\sigma\cdot n}{2}.
\label{eta-projectors}
\ee
Using them, one can expand the continuum spinor as
\be\label{ud-decomp}
\chi_I(\Omega,\tau)
=
\eta_+(\Omega)u_I(\Omega,\tau)
+
\eta_-(\Omega)d_I^\dagger(\Omega,\tau).
\ee
The use of $d^\dagger$ implements a particle-hole transformation in the
negative-energy band since
\be
\chi_I^\dagger(\sigma\cdot n)\chi_I =
u_I^\dagger u_I+d_I^\dagger d_I,
\label{normal-order}
\ee
up to a normal ordered constant.
Thus both $u$ particles and $d$ holes have the same positive eigenvalue
of $1$ at leading large $N$.
We note that under the global fermion-number symmetry
\be
\chi\longrightarrow e^{i\alpha}\chi,
\ee
one has
\be
\text{$u$ particle}:\quad u\longrightarrow e^{i\alpha}u,
\qquad
\text{$d$ hole}:\quad d\longrightarrow e^{-i\alpha}d
\ee
and so $u_I, d_I$ has equal and opposite charges.
Now using the orthonormality of $\eta_\pm$,
we obtain from \eq{ud-decomp} the action for the partons
\be \label{SF2}
S_F =
\int\dd\tau\,\dd\S\,
\; \left[ \sum_{I=1}^{N} \left(iu_I^\dagger\partial_\tau u_I
+i d_I^\dagger\partial_\tau d_I \right)
- j_A b_A - n_P \mu 
\right],
\ee
where $n_P$, $\rho$ and $j^A$ can be 
written in terms of the $u_I, d_I$'s as
\be \label{n-rho-j}
n_P =\sum_{I=1}^{N} (u_I^\dag u_I + d^\dag_I d_I),
\qquad \rho = q \sum_{I=1}^{N} (u_I^\dag u_I - d^\dag_I d_I),
\qquad j^A =  q \sum_{I=1}^{N} (
\zeta^A u_I^\dagger d_I^\dagger
+\bar\zeta^A d_Iu_I).
\ee
Here
\be
\zeta^A(\Omega)
:=\eta_+^\dagger\sigma_T^A\eta_-,
\qquad
\bar\zeta^A(\Omega)
:=\eta_-^\dagger\sigma_T^A\eta_+, 
\ee
are the off-diagonal matrix element of the tangential Pauli matrix
$\sigma_T^A := \s_a K_a{}^A$ between
the two local radial-spin states $\eta_+$ and $\eta_-$.
Note that the diagonal terms vanish:
$ \eta_\pm^\dagger\sigma_T^A\eta_\pm =0$.
For the northern-patch spinors \eq{eta_pm}, we obtain the explicit form
\be
\zeta^A = -i e^{-i\phi}(e^A_{\hat{\th}} -i e^A_{\hat{\phi}}).
\ee
We also note the useful identities
\be
\zeta^A \zeta_A =0, \qquad \zeta^A \bar{\zeta}_A =2
\ee
and
\be
\zeta^A \bar{\zeta}^B = h^{AB} + i \e^{AB},
\ee
where $h_{AB}$ is the metric for the 
unit-sphere 
\be
\dd\Omega_2^2
=h_{AB}\,\dd\Omega^A\dd\Omega^B
=\dd\theta^2+\sin^2\theta\,\dd\phi^2
\ee
and $\e_{AB}$ is the  covariant Levi–Civita tensors
\be
 \e_{AB}=\sqrt h\,\tilde{\e}_{AB},
 \qquad
 \e^{AB}=\frac{1}{\sqrt h}\,\tilde{\e}^{AB},
 \qquad \tilde{\e}^{\theta\phi}=+1.
\ee

\subsection{Guiding center dynamics}

It is well known that in the presence of a magnetic field,
a charged particle undergo a Hall drift. This happens the same for the
charged partons on the horizon. For a neutral black hole, there is no
net Hall current.  Let us derive the chiral Hall response of the
horizon of a charged back hole.

The Hall drift can be derived most simply
by considering  an individual parton and then sum over the electric
response. 
Consider a single parton state
\be \label{coh}
\ket{\psi(t)} = u(t) \ket{\O_+} + d(t)^\dag \ket{\O_-} ,
\ee
which describes a single parton ($u$ or $d$)
LLL wave packet localized near the point $\O $ on the fuzzy sphere. 
Physically, $\O_\pm$ is the position of the guiding center of the LLL orbital
for the $u$ or $d$ partons.
Dropping the filled sea constant, we obtain for  occupied levels
$u^\dag u =1$, $d^\dag d =1$: 
\be
L_K = i u^{\dag} \dot{u}  +\cA_{A} \dot{\xi}^A_+
+ i d^{\dag} \dot{d} - \cA_{A} \dot{\xi}_-^A,
\ee
where $-i \bra{\O_\pm} d \ket{\O_\pm} = \cA_A(\O_\pm) d\xi_\pm^A$ is
the Berry monopole
connection and $\xi_{(\pm)}^A$, $A=1,2$ are the guiding center
coordinates of the $u$ and $d$ on the unit sphere.
Including the flavor, we obtain
\be \label{LK1}
L_K =  \sum_{p\in \cS}i \eta_p^{\dag} \dot{\eta}_p  \pm \cA_A(\O_p) \dot{\xi}_p^A,
\ee
where $\eta = u$ or $d$, the $+$ (resp. $-$)
sign is chosen for the $u$ (resp. $d$) -partons. and
$\cS$ denotes  the set of occupied partonic LLL state.
For example, for the half Fermi sea described by \eq{rsNQ},
there are $r$ of $u$-partons and $s$ of $d$-partons. 
We note that the
$u$ and $d$ partons see opposite Berry curvature. This will lead to
a nontrivial Hall response of the fuzzy 
horizon when the black hole is charged.
%% We  note also
%% that one may consider  state of more general form than \eq{coh}. For
%% example,
%% \be
%% \ket{u(t)} = \sum_m \eta_m \varphi_m \ket{J,m},\quad \varphi_m := \la J,m| \O\ra
%% \ee
%% for the $u$-part.
%% Here  $\eta_m$ describes
%% the occupation amplitudes of the $m^{th}$ LLL orbitals and
%% does not describe a localized particle on the sphere. In the special case of
%% $\eta_m = \eta$ for all $m = -J, \cdots, J$, we obtain the one particle
%% wave-packet \eq{coh} which describes  the classical
%% guiding center
%% dynamics of a localized LLL parton.

It is interesting to note that the usual kinetic energy is quenched and a
first order Lagrangian is obtained for these horizon partons. 
The situation resembles the quantization of open string in a
constant NS-NS $B$-field \cite{Chu:1998qz} when the field is large.
There a Moyal type 
non-commutative geometry is obtained on the worldvolume of a D-brane.
Here, the
partons' guiding center coordinates obey the noncommutative geometry
(see appendix)
\be \label{xxx}
[x_a,x_b] =\pm i \e_{abc} x_c, \quad a,b,c, = 1,2,3
\ee
where $x_a = J n_a$ is the Cartesian coordinates of the partons and the
$\pm$ sign is for the $u/d$ partons. 
 It is intriguing that the endpoints of
 open string also see an opposite sign of noncommutative geometry
 \cite{Chu:1998qz}
which is due to the fact that the endpoints of open-string are
oppositely charged. The close analogy of our result with string
quantization suggests that our
black hole quantum mechanics may have a stringy origin.
It is remarkable that the
noncommutativity in this system is not merely the geometric
noncommutativity of the fuzzy sphere. The horizon partons themselves
exhibit intrinsically noncommutative dynamics, arising from the Berry
monopole (B)-field on the horizon. Thus the noncommutative structure
is both kinematical, encoded in the fuzzy-sphere geometry, and
dynamical, encoded in the Lowest-Landau-Level motion of the partons.

\subsection{Hall current and Ohmic current on fuzzy horizon}

So far the LLL is derived for a pure fuzzy sphere background. Let us next
consider a perturbation \eq{XqB}
to include a
horizon gauge field $b_A$ on the fuzzy sphere.
Consider a single LLL
parton, the corresponding charge and guiding center current is given by
\be \label{jj}
\r(\t,\O) = \pm q \frac{\d^{(2)}(\O - \xi(\t))}{\sqrt{g}},\quad
j^A =  \pm q \dot{y}^A \frac{\d^{(2)}(\O - \xi(\t))}{\sqrt{g}},
\ee
where the sign $+/-$ is for $u/d$ and where $d y^A = R d \xi^A$ is the
displacement vector on  $S^2_R$.
Substituting into \eq{SF1},
we obtain the action for the LLL parton
\be
S_{LLL} = \pm \int d\t \; (
\cA_A \dot{\xi}^A - q R b_A \dot{\xi}^A \mp \mu).
\ee
Ignoring the chemical potential, we obtain the equation of motion for
each LLL parton,
\be \label{hall-drift}
(\cF_{AB} -q F_{AB}) \dot{y}^B = -q E_A,
\ee
where
\be
E_A = - \del_\t b_A
\ee
is the physical electric field on the sphere $S^2_R$
and 
\be
\cF = \frac{J}{R^2} E^1 E^2
\ee
is the monopole field strength with respect to the vielbein $E^1 = R d \th,
E^2 = R \sin \th d \phi$. 
As the monopole field strength is of
order $N$, we can  treat the external $B$-field as  a perturbation
whenever it is small compared to the monopole. In case $\cF_{AB} \gg
f_{AB}$, we
obtain the Hall drift
\be \label{ydot}
\dot{y}^A = q (\cF^{-1})^{AB} E_B.
\ee
The result
tells us that the charge distribution localized around the
guiding center drifts in the perpendicular direction to an applied external
electric field.
We remark that the  field theory current operator \eq{n-rho-j} is
off-diagonal in the local radial-spin basis while the guiding current
in \eq{jj} is diagonal. It is interesting to note how \eq{jj} can
be obtained from \eq{n-rho-j}. 
Physically, what happens is that
an applied tangential electric field, when
represented by a time-dependent $b_A$ in the temporal gauge,
induces virtual mixing between the two radial-spin sectors due to
the interaction
\be
H_{\rm int} = - j^A b_A = -  q ( \zeta^A u^\dagger d^\dagger
+\bar\zeta^A d u ). 
\ee
This should not be interpreted as ordinary spin propagation
on the fuzzy sphere.  After projecting to the LLL, the
same response is equivalently described by guiding-center
Hall drift within the projected Hilbert space and the guiding
center current \eq{jj} is obtained.

Let us derive the Hall conductance of the Fermi sea.
Note that the drift velocity is the same for the $u$ and oppositely charged
$d$-partons. Therefore we obtain the total current for the filled LLL,
\be
j^A = (\r_u -\r_d)q \dot{y}^A,
\ee
where $\r_u = r/(4\pi R^2)$,   $\r_d = s/(4\pi R^2)$ are the number density
for the partons and $r,s$ does not need to be equal
for a charged black hole \eq{rsNQ}. As a result we obtain
\be
j^A = \e^{AB} \s_H E_B , \qquad
\s_H :=  (\r_u -\r_d)\frac{q^2 R^2}{J}, 
\ee
where $\s_H$ is the Hall conductance and $\e^{12} =+1$ is for
the orthonormal frame.
Using the  filling fraction $\n_s:=N_s/N^2$ for the species $s=u,d$,
one can also write $\s_H$ as
\be \label{ideal}
\s_H = \frac{Nq^2}{2\pi} (\n_u-\n_d).
\ee
%%c9
%% Here $\nu_s=N_s/N^2$ is the occupancy fraction averaged over
%% the $N$ flavor copies and the $N$ orbital states of each LLL.
%% The corresponding single-flavor Hall filling is $N_s/N$, so the
%% standard formula $\sigma_{xy}=\nu q^2/(2\pi)$ gives the extra factor
%% of $N$ in \eq{ideal}.
One can also write $\s_H$ in terms of charge of the horizon,
\be
\s_H = \frac{q Q}{2\pi N}.
\ee
It is intriguing that the quantum
horizon has a Hall response
when the black hole is charged.

The above derivation is based on the single-particle
dynamics of the partons
and gives a nondissipative Hall current.
 A black-hole horizon, however, acts as an effective thermal
 environment, and each parton interacts with the remaining horizon
 degrees of freedom as with a many-body bath. In fact, in the quantum
 mechanics, the fermion two-point function exhibits decay with a
 characteristic rate $\Gamma_d\sim R^{-1}, R=Nl_P$, or,
 equivalently, over a relaxation time $t_d\sim R$. This scaling is
 consistent with the Hawking temperature $T_H\sim R^{-1}$, and
 therefore supports the interpretation of the horizon as a thermal
 bath. However, the decay of the correlator alone does not prove exact
 thermality. A rigorous analysis would require  verifying the
 Kubo–Martin–Schwinger condition and the associated
 fluctuation–dissipation relation. We leave such an analysis for future
 work.
For present purpose, we model the effective dynamics of the partons
with the first-order Langevin equation of the form
\be
  \cF_s \e_{AB}v_s^{B}
  = - q_s E_{A}  +\eta v_{sA},
  \label{Langevin}
\ee
where the indices refer to the orthonormal frame $E^A$
and $\eta$ is a friction coefficient.
Here $s=u,d$ denotes the species of partons, $q_s = \pm q$,
$\cF_s = \pm \cF$ with $\cF= J/R^2$ is the
Berry field strength,  and $\pm$ applies
for $u/d$.
Solving \eq{Langevin}, we obtain
\be
  v_s^{A}
  =\frac{q_s}{\eta^2+\cF^2}
  \left(
    \eta\delta^{AB}
    +\cF_s\e^{AB}
  \right)E_{B} .
  \label{velocity_solution}
\ee
The current from sector $s$ is
\be
  j_s^{A}=\rho_s q_s v_s^{A},
\ee
where 
\be
\r_s = \frac{N_s}{4 \pi R^2}, \qquad N_u =r, \quad N_d =s, 
\ee
is the number density of the species $s$.
Introducing the dimensionless parameter $\gamma_T := \eta/\cF$
to characterize the dissipation effect. We have
\be
  j_s^{A}
  =\sigma_{xx}^{(s)}E^{A}
  +\sigma_H^{(s)}\e^{AB}E_{B},
\ee
where
\be  \label{sigma_s}
  \sigma_{xx}^{(s)}
  :=\frac{N q^2\nu_s}{2\pi}
  \frac{\gamma_T}{1+\gamma_T^2}, \qquad
  \sigma_H^{(s)}
  :=\frac{N q^2 \nu_s}{2\pi}
  {\rm sign}(q_s)
  \frac{1}{1+\gamma_T^2} .
\ee
Note that
the longitudinal conductivities of the two sectors add, while their Hall
conductivities have opposite signs. Summing the two sectors, the total
current is given by a Ohmic part and a Hall part:
\be \label{jA}
  j^{A}
  =\sigma_{xx}E^{A}
  +\sigma_H\e^{AB}E_{B},
\ee
where
\be \label{sigmaxx_equalgamma}
  \sigma_{xx}
  =\frac{Nq^2}{2\pi}(\nu_u+\nu_d)
  \frac{\gamma_T}{1+\gamma_T^2} , \qquad
  \sigma_H
  =\frac{Nq^2}{2\pi}(\nu_u-\nu_d)
  \frac{1}{1+\gamma_T^2} .
\ee
%c13
The transport law \eqref{sigmaxx_equalgamma} assumes that the applied field
does not create real horizon partons. The fixed-parton-number,
one-body guiding-center description is therefore controlled for
$\hat\omega\ll\epsilon_0$, or
\be
 \hat\omega l_P \ll 1.
 \label{fixed_parton_condition}
 \ee
 Since the  $U(1)$ charge is
 conserved, the first real excitation is a $u-d$ pair with threshold
$\sim 2 \e_0$. 
 Once $\hat\omega$ reaches  this threshold,
 the field can populate sectors with different
total parton number and the fixed-parton transport description must be
replaced by a scattering matrix of black-hole.

A simple thermal estimate gives $\eta \sim T^2$. In fact
the friction coefficient is defined by 
dissipated power $P_{\rm diss} = \eta v$.
For collisions on the thermal horizon,
if each dissipative event releases an energy of order $T$ and since the
relaxation rate is $\t_d \sim T$ , then 
the power loss scale as $P_{\rm diss} \sim \t_d \D E  \sim T^2$ and hence
 $\eta \sim T^2 $. 
Since $\cF \sim 1/(N l_P^2)$, this gives
\be
\g_T = \frac{c_T}{N}
\ee
for some $N$-independent constant $c_T$. As a result, an order-one
Ohmic conductivity is obtained:
\be
  \sigma_{xx}
  =\frac{q^2c_T}{2\pi}(\nu_u+\nu_d)
  +\cO(N^{-2}) .
  \label{ohm_scaling}
\ee
For $\nu_u+\nu_d = 1$  at half filling, we have
\be
  \sigma_{xx}\simeq \frac{q^2c_T}{2\pi} 
\ee
and is of order 1 independent of $N$.
Note that the precise value of $c_T$ is phenomenological in this
Langevin treatment, however the large-$N$ scaling is robust. So we
get an Ohm's law for black hole with a Ohmic conductivity that is 
independent of black hole.
Microscopically, the frictional coefficient can be computed from the
retarded correlator of the frictional force via a Kubo formula.
We conjecture that this results in $c_T =1/(2q^2)$
and so  the membrane-paradigm value of
$1/4\pi$ is recovered. 
As for the Hall conductivity, we have
\be
  \sigma_H
  =\frac{Nq^2}{2\pi}(\nu_u-\nu_d)
  \left[1-\frac{c_T^2}{N^2}+\cO(N^{-4})\right] .
  \label{hall_scaling}
\ee
Thus the thermal friction changes the Hall response
only at order $1/N^2$ relative to the ideal LLL value \eq{ideal}.

\section{Interacting Multi-Blocks Matrix Quantum Mechanics}

Next, let us discuss the interaction of systems
described by matrix blocks in the matrix quantum mechanics.
Consider a system of $\cK$ isolated systems of quantum space bits
described by
matrices $X^{(i)}_a$ of rank $N_i$,
each described by  Lagrangian  of the form
\be
L^{(N)}[X,\psi]
=
\Tr_{N}\!\left[
\frac{1}{2 M_0}\,\dot X_a^2
+
\l_{N}\left([X_a,X_b]^2+4X_aX_a\right)
+
i \dot\psi^{\dagger}\psi
- a_2 \l_{N} \,\psi^{\dagger}\sigma^aX_a\psi
\right],
\label{ChuExact}
\ee
where
\be
M_0=a_0^2\MP,
\qquad
\lambda_N:=\frac{\MP}{N^2}.
\label{nodecoeff}
\ee
We have ignored
the noncommutative energy term $r_X$ here.
To introduce interaction between them in matrix model, one 
introduces off-diagonal $N_i\times N_j$ link variables  $W^{ij}_a$.

\subsection{Two interacting blocks}

Let the two node systems have ranks $N_1, N_2$. To describe their
interaction
in matrix model, let us package them as the Hermitian block
matrices: 
\be
\mathbf X_a
=
\begin{pmatrix}
X_a& W_a\\
W_a^\dagger&Y_a
\end{pmatrix},
\qquad X_a\in\operatorname{Mat}_{N_1},
\quad
Y_a\in\operatorname{Mat}_{N_2},
\quad
W_a\in\operatorname{Mat}_{N_1\times N_2}
\label{twoXblock}
\ee
for the bosonic variables, 
and
\be
\mathbf{\Psi}
=
\begin{pmatrix}
\psi_1&\eta^{12}\\
\eta^{21}&\psi_2
\end{pmatrix}, \qquad
\eta^{12}\in\operatorname{Mat}_{N_1\times N_2},
\quad
\eta^{21}\in\operatorname{Mat}_{N_2\times N_1}
\label{twoPsiblock}
\ee
for the fermionic variables. 
Here $\eta^{12}$ and $\eta^{21}$ are
independent oriented complex Grassmannian links.
Unlike the bosonic links, one does not impose $\eta^{21}=\eta^{12\dagger}$.
Their Hermitian conjugates appear in $\mathbf\Psi^\dagger$.
It is convenient to introduce 
the canonically normalized link variables $Z_a$ defined by
\be
W_a=\sqrt{M_0}\,Z_a=a_0\sqrt{\MP}\,Z_a,
\label{canonZ}
\ee
so that its kinetic term is $\Tr\dot Z_a^\dagger\dot Z_a$.
On the block configurations, the theory has the residual global symmetry 
\begin{align}
  & X_a \longmapsto U_1X_aU_1^\dagger, \qquad
  Y_a \longmapsto U_2Y_aU_2^\dagger, \qquad
  Z_a \longmapsto U_1Z_aU_2^\dagger, \label{res-symm1}\\
  &  \psi_i \longmapsto U_i\psi_i, \qquad
  \eta^{ij} \longmapsto U_i\eta^{ij}U_j^\dagger,\label{res-symm2}
\end{align}
where the links $Z_a, \eta^{ij}$ are bi-fundamental variables.

We propose the interacting model to be given by the following
projector-weighted quantum mechanics
\be
\begin{aligned}
L_{12}=
\frac{1}{2M_0}\Tr\dot{\mathbf X}^2_a
+
\Tr\!\left[
\widehat\lambda_{12}
\left([\mathbf X_a,\mathbf X_b]^2+4\mathbf X_a\mathbf X_a\right)
\right] +
i \Tr\dot{\mathbf\Psi}^{\dagger}\mathbf\Psi
-
a_2\Tr\!\left(
\widehat\lambda_{12}\,
\mathbf\Psi^{\dagger}\sigma^a\mathbf X_a\mathbf\Psi
\right),
\end{aligned}
\label{compact2}
\ee
where
\be
\widehat\lambda_{12}=:
\lambda_1P_1+\lambda_2P_2,
\qquad P_1=
\begin{pmatrix}\mathbf1_{N_1}&0\\0&0\end{pmatrix},
\quad
P_2= \begin{pmatrix}0&0\\0&\mathbf1_{N_2}\end{pmatrix},\quad
\lambda_i=\frac{\MP}{N_i^2}.
\ee
We note that when the link
variables are turned off: $Z_a=\eta^{12}=\eta^{21}=0$,
we have
\be
L
=
L^{(N_1)}[X,\psi_1]
+
L^{(N_2)}[Y,\psi_2]
\ee
and correctly reproduces the physics of the two isolated systems.
To explicitly
display the interaction terms described by the link variables, let us introduce
\begin{align}
&F^X_{ab} :=[X_a,X_b], \quad F^Y_{ab} :=[Y_a,Y_b], \label{qq1}\\
&h_{ab}:=\mathcal D_aZ_b-\mathcal D_bZ_a,
\quad\mbox{where}\quad \mathcal D_aZ_b:=X_aZ_b-Z_bY_a, \label{qq2}\\
& q^X_{ab} :=Z_aZ_b^\dagger-Z_bZ_a^\dagger, \quad
q^Y_{ab} :=Z_a^\dagger Z_b-Z_b^\dagger Z_a. \label{qq3}
\end{align}
We have the link part of the bosonic Lagrangian:
\be
\begin{aligned}
L^{(12)}_{B,\mathrm{link}}={}&
\Tr\dot Z_a^\dagger\dot Z_a
+
4M_0(\lambda_1+\lambda_2)\Tr Z_a^\dagger Z_a
\\
&-
M_0(\lambda_1+\lambda_2)
\Tr(h_{ab}h_{ab}^\dagger) +
2M_0\left[
\lambda_1\Tr_{N_1}(F^X_{ab}q^X_{ab})
+
\lambda_2\Tr_{N_2}(F^Y_{ab}q^Y_{ab})
\right]
\\
&+
M_0^2\left[
\lambda_1\Tr_{N_1}(q^X_{ab})^2
+
\lambda_2\Tr_{N_2}(q^Y_{ab})^2
\right].
\end{aligned}
\label{boslink2}
\ee
Here the first line describes the free part of the links, the second
describe their quadratic interaction with the background, and the third line
gives their self-interaction.
We note that the mass term has a
nontrivial coefficient
\be
4M_0(\lambda_1+\lambda_2)
=
4a_0^2\MP^2\left(\frac1{N_1^2}+\frac1{N_2^2}\right).
\ee
The sign is positive and signals a tachyonic instability at the origin.
As we will show below, it is remarkable that the $Z$'s is stabilized as a result
of the interacting terms. This Higgs mechanism in the matrix model locks the
horizon gauge field with the bulk Maxwell field and 
is vital to the emergence of the horizon boundary condition and a membrane
description. 

We will not discuss much about the fermionic part.
But for completeness, it is
given here
\be
\begin{aligned}
L^{(12)}_{F,\mathrm{link}}={}&
i\Tr_{N_2}\!\left(\dot\eta^{12\dagger}\eta^{12}\right)
+
i\Tr_{N_1}\!\left(\dot\eta^{21\dagger}\eta^{21}\right)
\\
&-
a_2\lambda_1\Tr_{N_1}\!\left[
\eta^{21\dagger}\sigma^aY_a\eta^{21}
+
\sqrt{M_0}\left(
\psi_1^\dagger\sigma^aZ_a\eta^{21}
+
\eta^{21\dagger}\sigma^aZ_a^\dagger\psi_1
\right)
\right]
\\
&-
a_2\lambda_2\Tr_{N_2}\!\left[
\eta^{12\dagger}\sigma^aX_a\eta^{12}
+
\sqrt{M_0}\left(
\eta^{12\dagger}\sigma^aZ_a\psi_2
+
\psi_2^\dagger\sigma^aZ_a^\dagger\eta^{12}
\right)
\right].
\end{aligned}
\label{fermlink2}
\ee
The complete two-block theory can therefore be written as
\be
\begin{aligned}
L={}&
L^{(N_1)}[X,\psi_1]
+
L^{(N_2)}[Y,\psi_2]
+
L^{(12)}_{B, \mathrm{link}}[Z,X,Y]
+
L^{(12)}_{F,\mathrm{link}}[\eta^{12},\eta^{21}, Z,  X, Y, \psi_1,\psi_2] \;.
\end{aligned}
\label{full2expanded}
\ee

We remark that 
one may instead retain only the original node fermions by imposing
\be
\eta^{12}=\eta^{21}=0.
\ee
In this case we obtain a {\it minimal model} with
$L^{(12)}_{F, \mathrm{link}}=0$ 
and the fermions of the two systems interact only indirectly through
the dynamical bosonic blocks.

\noindent\underline{FIDO frame}
 
For the application of black hole ($N_1=N$) interaction with an
environment block ($N_2$) . We
can use the FIDO time determined from the $N$-block,
\be
\tau = \frac{2a_0 M_P l_P}{N} t.
\ee
In this case,  the action for each block is
\be
S_i[X,\psi] = \frac{l_P}{Na_0}\int d\t 
\Tr_{N_i}\!\left[\dot X_a^2
+
\a_{i}\left([X_a,X_b]^2+4X_a^2\right) \right]
+
\int d\t \Tr_{N_i} \left[ i \dot\psi^{\dagger}\psi
- \frac{a_2 l_P}{a_0 N} \a_i \,\psi^{\dagger}\sigma^aX_a\psi
\right], \nn
\ee
where
\be
\a_i := \frac{N^2}{2 N_i^2 l_P^2}.
\ee
For the link part of the bosonic action, we find
it more convenient to use the original \eq{canonZ}
link fields $W_a$ instead of the rescaled $Z_a$'s. 
The result is
\bea
S_{B, {\rm link}}^{(12)} 
= && \frac{l_P}{a_0N_1}\int \dd \t  
\Big[ 2 \Tr_{N_1} (\dot W_a\dot W_a^\dagger)
+ 4 (\a_1+ \a_2)\Tr_{N_1} (W_aW_a^\dagger) \nn\\
&& -(\a_1+ \a_2)\Tr_{N_1}(h_{ab}h_{ab}^\dagger)
+
2\a_1\Tr_{N_1}(F^X_{ab}q^X_{ab})
+ 2\a_2 \Tr_{N_2}(F^Y_{ab}q^Y_{ab}) \nn\\
&&  +  \a_1\Tr_{N_1}(q^X_{ab})^2
    +   \a_2 \Tr_{N_2}(q^Y_{ab})^2 \Big] .
\label{quadratic-link}
\eea
where $h_{ab}, q_{ab}^X, q_{ab}^Y$ are defined as in \eq{qq2} ,\eq{qq3}
but with $Z$
replaced by $W$, that is explicitly,
\begin{align}
&h_{ab}:=\mathcal D_aW_b-\mathcal D_bW_a,
\quad\mbox{where}\quad \mathcal D_aW_b:=X_aW_b-W_bY_a, \label{qq2-W}\\
& q^X_{ab} :=W_aW_b^\dagger-W_bW_a^\dagger, \quad
q^Y_{ab} :=W_a^\dagger W_b-W_b^\dagger W_a. \label{qq3-W}
\end{align}
As for the fermions, we have the link part
\bea
S_{F,{\rm link}}^{(12)}= &&
\int \dd \t \;\; i\Tr_{N_2}(\eta_{12}^\dagger\dot\eta_{12})
+i\Tr_{N_1}(\eta_{21}^\dagger\dot\eta_{21})\nn \\
&&-\frac{a_2 l_P}{a_0N}
\alpha_1\Tr_{N_1}\left[
\psi_X^\dagger\sigma^a\left(X_a\psi_X+W_a\eta_{21}\right)
+\eta_{21}^\dagger\sigma^a\left(W_a^\dagger\psi_X+Y_a\eta_{21}\right)
\right]
\nn \\
&& -\frac{a_2 l_P}{a_0N}\alpha_2\Tr_{N_2}\left[
\eta_{12}^\dagger\sigma^a\left(X_a\eta_{12}+W_a\psi_Y\right)
+\psi_Y^\dagger\sigma^a\left(W_a^\dagger\eta_{12}+Y_a\psi_Y\right)
\right] . 
\label{fermion-kinetic}
\eea

\subsection{Gauss law constraint}
\label{sub-Gc}

Before we move on, let us make an important remark concerning the status
of $SU(N)$ symmetry in our model.
In the original formulation \cite{Chu:2024qil} of
the matrix model, there is no $X_0$ and  an ordinary time derivative
is featured. As such the model can be understood as a model with a global
$SU(N)$ or as a $A_0=0$ temporal gauge fixed version of a local $SU(N)$
theory.
For a discussion of the difference between treating the
$SU(N)$ symmetry as a global or a local  symmetry in   the BFSS matrix
model, see \cite{Maldacena:2018vsr}.

As we discussed above, it is necessary to consider the matrix model action
\eq{L}, or its multiple block extension \eq{compact2}, \eq{compactK}
as a temporal gauge $A_0=0$ fixed theory of the covariantized theory with
the time derivatives replaced by the covariant derivatives
\be
D_\tau \mathbf X_a =  \dot{\mathbf X}_a - i [A_0,  \mathbf X_a],
\qquad
D_\tau \mathbf \Psi = \dot{\mathbf \Psi} - i A_0 \mathbf \Psi.
\ee
As a result, the analysis of the gauge fixed theory should be complemented
with the imposition of the Gauss law constraint.
In this subsection, we show that imposing the Gauss law constraint
leads to the important conclusion that the net charges of the
partons on the horizon  can be identified with the EM
charge of black hole.

The reason is clear.  As the black hole horizon
block is part of a bigger system together with an environmental block
and the associated link variables
\be
\mathbf X_a
=
\begin{pmatrix}
X^H_a& W_a\\
W_a^\dagger&X^E_a
\end{pmatrix},
\qquad
\mathbf{\Psi}
=
\begin{pmatrix}
\psi^H&\eta\\
\tilde{\eta}&\psi^E
\end{pmatrix},
\ee
Varying $A_0$ gives the total Gauss law constraint of the form
\be \label{cGc}
G = i[\mathbf X_a, \mathbf P_a] +\rho_\mathbf{\Psi} =0.
\ee
Here $ \mathbf P_a$ is the momentum conjugate to 
$\mathbf X_a$ and $\rho_\Psi$ 
is the matrix charge density carried by the fermions.
After the constraint \eq{cGc} is derived, one may set $A_0 =0$ and
consider the gauge-fixed action, together with
the Gauss law constraint on physical states.
Note that the Gauss law is imposed on the total
black-hole-plus-environment system, not separately on the
black-hole block.
In block form, the diagonal parts of the constraint are schematically
\begin{align}
  G_H &= i[X_a^H,P_a^H]  + \rho_W^H + \rho_\psi^H =0,
  \label{GH}\\[2mm]
  G_E &= i[X_a^E,P_a^E] + \rho_W^E + \rho_\psi^E =0,
  \label{GE}
\end{align}
where the link-field charge densities are
\be
  \rho_W^H = i(W_a\Pi_a^\dagger-\Pi_a W_a^\dagger),
  \qquad
  \rho_W^E =  - i(W_a^\dag \Pi_a - \Pi^\dag_a W_a) 
  \ee
  and
  \be
  \rho_\psi^H = \psi_H \psi_H^\dag+\eta \eta^\dag,
  \qquad
  \rho_\psi^E = \psi_E \psi_E^\dag + \tilde{\eta} \tilde{\eta}^\dag.
  \ee
Note that 
\be \label{W0}
\Tr_H \rho_W^H = - \Tr_E \rho_W^E 
\ee
Physically, the link field
$W_a$ is bi-fundamental: it carries one unit of charge under the
$H$-block gauge symmetry and the opposite charge under the 
$E$-block gauge symmetry. Therefore it can transfer charge
between the black-hole block and the environment, but it does not
create net charge in the total two-block system.
Now, in the continuum fuzzy-sphere limit the commutator
term in \eqref{GH} becomes the intrinsic divergence
of the horizon electric field,
$ i[X_a^H,P_a^H] \;\longrightarrow\; D^A E^H_A$, 
thus the local horizon constraint takes the form
\be
  D^AE^H_A = \rho_\psi^H+\rho_W^H .
\ee
Integrating over the closed sphere removes the divergence term
$  \int_{S^2} d\Sigma\, D^AE^H_A =0$ 
and gives only the zero-mode constraint
\be  \label{psi0}
\Tr_H  \rho_\psi^H  +\Tr_H \rho_W^H=0 .
\ee
On the other hand,
the environmental constraint supplies the radial electric flux.
In the continuum
limit, the link charge appears as a surface source localized at the
stretched horizon,
so that
\be
  \nabla_i E^i = \rho_W^E + \rho_\psi^E .
\ee
Integration this in the region external to the
stretched horizon and since the
charge of the black hole can be obtained from the flux, we have
\be
Q_{BH} = \int_{S_R^2} d\Sigma\, E_r =  \Tr \mathbf{\Psi}\mathbf{\Psi}^\dag,
  \label{bulkzero}
  \ee
  where we have used \eq{W0} and \eq{psi0}.
  In the low-energy regime with energy small compared to the
  condensation scale, $\psi_E$ and the off-diagonal fermionic links
$\eta$, $\tilde{\eta}$ are not excited so that 
  the charge is carried entirely by the horizon states $\psi_H$.
  After the usual particle–antiparticle mode decomposition
and normal ordering, their contribution is
\be
Q_{\rm BH}=q(r-s),
\ee
where $r$ and $s$ are the occupation numbers of the $u$ and $d$ sectors.
In this case, the Gauss law identifies the asymptotic electric
flux directly with the net horizon parton charge.

\section{Locking Mechanism and Quantum Membrane Paradigm}

Let us now discuss the matrix interaction involving
a black hole block described by a $N\times N$ block and an environmental
block described by a $N' \times N'$ Cartesian block. 

The black-hole block is expanded around a spin-$J$ fuzzy sphere:
\be
  X_a^{\HH}=J_a+ R B_a,
 \qquad
 [J_a,J_b]=\ii\epsilon_{abc}J_c,
 \quad
 J_a^2 =J(J+1)\one_N,\quad J=\frac{N-1}{2}. 
 \label{BHbackground}
\ee
At large $N$, $R =  N l_P$ and the horizon fluctuation decomposes as
\be
 B_a(\Omega)=K_a{}^A(\Omega)b_A(\Omega)+n_a(\Omega)\upvarphi_{\HH}(\Omega),
 \label{Bdecomp}
\ee
where $K_a{}^A$ is the Killing tangent frame, $n_a$ is the outward
unit normal, $b_A$ is the tangential horizon gauge field, and
$\upvarphi_{\HH}$ is the normal scalar.
Using the FIDO time of the black hole block,
the  action takes the Maxwell-scalar form
\be \label{SH2'}
S_\HH = \frac{R^2}{2 g_{\rm BH}^2}\int d\t d\O
\left[
  (\del_\tau b_A)^2 + (\del_\tau  \upvarphi_{\rm H})^2 - \frac{1}{l_P^2}
  \left(f^2+ (D_A  \upvarphi_{\rm H})^2 +4 \upvarphi_{\rm H}^2
- 8 f \upvarphi_{\rm H}\right)
\right],
\ee
where 
\be
\frac{1}{2 g_{\rm BH}^2} = \frac{l_P}{4\pi a_0}.
\ee

The environmental block of rank $N'$ is written as
\be
X_a^{\EE}=LY_a,
\ee
with
\be
L = \frac{N'  l_P}{N}, \qquad
Y_a = y_a +  \frac{R}{L}A_a(x) , \qquad
y_a = \frac{x_a}{2 l_P L} - i \frac{\del}{\del x_a},
\ee
where $x_a$ plays the role of the coordinates for
the emerged 3 dimensional space. 
With the trace prescription
\be \label{envtrace}
 \frac{1}{N'}\Tr_{N'}\longrightarrow\frac{1}{V_3}\int\dd^3x
 \ee
and choose the regularized volume of space to be $V_3 = 24 R^2 L$, 
we obtain in the bulk the Maxwell action in Gauss unit
\be\label{Maxwell}
 S_{\rm env}
 =\frac{1}{8 \pi}
 \int\dd\tau\,\dd^3x
 \left[(\partial_\tau A_a)^2-\frac{1}{2} F_{ab}^2 \right].
\ee
Near the horizon, we will decompose $A_a$ in accordance with the
geometry of $S^2$ as
\be
 A_a=K_a{}^Aa_A+n_a\upvarphi_{\EE}.
 \label{Adecomp}
 \ee
 
 We will now show that in a thin region near the horizon,
 the gauge field $b_A$ on the fuzzy sphere is locked to the
 bulk gauge field $a_A$ due to a novel mechanism. As a result, the
 partonic LLL current on the membrane horizon responds to the bulk
 Maxwell field and becomes physical. 
 To explain the locking, let us
 combine the black hole block and the environmental block
 into an interacting system with
\be
 \mathbb X_a=
 \begin{pmatrix}
 X_a^{\HH}&W_a\\
 W_a^\dagger&X_a^{\EE}
 \end{pmatrix}.
 \label{blockmatrix}
\ee
Taking the limit of large $N'$, we obtain from
\eq{boslink2} the Lagrangian for the link variables
\be\label{L-link}
L_W = \frac{  l_P}{a_0N}\left[
  2 \Tr_\HH \dot W_a \dot W_a^\dagger -U_2 -U_4\right],
\ee
where 
\be
U_2 =\a \Tr_{\HH}
\left[ h_{ab}h_{ab}^\dagger - 2 F^{\HH}_{ab}q^{\HH}_{ab}
  -4 W_aW_a^\dagger \right], \quad
U_4 = - \a  \Tr_{\HH} \; q^{\HH}_{ab}{}^2, \quad \a := \frac{1}{2l_P^2}.
\ee
Here $U_2$ is quadratic in $W$, and  includes the
  mass term as well as the
  linear and quadratic interaction of $W$ with the
  gauge fields;
  and $U_4$  denotes the quartic self-interaction of the $W$'s.
  One may canonically normalize $W_a$ for the kinetic terms. However the
  potential energy is given by
  \be \label{VU24}
V= \frac{  l_P}{a_0N} (U_2+U_4)
  \ee
  in any case.
Note that   $W$ is bi-fundamental
with respect to the gauge fields and so it's 
covariant derivative takes the form
$ D_A W = \del_A W + i (b_A -a_A)W$.
As we will show below, the link potential is Higgs-like
  with a tachyonic mass term and a positive
  quartic
  interaction term. This induces a nonvanishing condensate and turns
  the kinetic term of the link
  into a potential term  for the relative gauge field:
  \be \label{VC0}
|D_A W|^2 \sim |W^2| (b_A -a_A)^2.
\ee
This results in the vacuum configuration
\be \label{lock-ba}
b_A = a_A
\ee
where the gauge fields are locked.
It turns out the potential is steep as $|Z^2|$ is large
and so the locking is effective.

\subsection{Condensate of links and locking of gauge fields}

Next let us use \eq{L-link} and give a full 
derivation of the locking.
%c5 by spelling out
% the explicit form of the covariant derivative coupling for  the $Z$,
% and demonstrating the existence of a $Z$-condensate.
To determine the link instability, we first  turn off the perturbation
$B_a =A_a =0$
and consider the link self-interaction.
With the background
\be\label{bkgd-XY}
  \bar X_a^H=J_a,
  \qquad
  \bar X_a^E=Ly_a,
\ee
we have
\be \label{FH}
  F^H_{ab}=i\epsilon_{abc}J_c,
  \qquad
  F^E_{ab}=0,
\ee
and so
\be \label{V2}
U_2 = \a \Tr_\HH \left[
  {h}_{ab}^\dag {h}_{ab}- 4i \e_{abc} J_c W_a W_b^\dagger  
- 4 W_a^\dagger W_a \right],
\ee 
where 
\be
h_{ab}= \bar{h}_{ab} :=   \bar{\cD}_a W_b - \bar{\cD}_b W_a,
\quad\mbox{with} \quad
\bar{\cD}_a W_b = (J_a - \xi_a) W_b, \quad 
\xi_a := L y_a
\ee
is the background derivative.
Let us denote
\be \label{xi-L-y}
\xi_a :=Ly_a= \frac{x_a}{2l_P} + L p_a
\ee
for
the value of the environmental background \eq{bkgd-XY}.
We can also write $\xi_a = \xi n_a$ in polar form
with $n_a$ parametrized by polar angles $\O$.
It is convenient to introduce the basis of ``position'' eigenstates
defined by
\be
\bar{X}^E_a \ket{\xi, \O} = \xi n_a \ket{\xi, \O}, \qquad 
\ee
Evaluating \eq{V2}, we obtain $U_2$ in the form
\be
U_2 = \a \Tr (W_a^\dag \cK_{ab} W_b),
\ee
where $\cK_{ab}$ is the quadratic operator
\be\label{cK}
\cK_{ab}(\xi)
  := \cH_{ab}(\xi)
  -4\delta_{ab}
  -4(\vect J\cdot\vect S)_{ab}.
\ee
Here
\be
\cH_{ab} :=  2\left(\bar{\cD}^2\delta_{ab}-\bar{\cD}_b\bar{\cD}_a\right)
\ee
is  the positive curl energy. The second term in \eq{cK}
arises from
the negative mass term of the model. The third term in \eq{cK}
arises from the coupling of the background \eq{FH} with the spin-one
generators $S_c$ that act on the vector index as
\be
  (S_c)_{ab}=-i\epsilon_{cab}.
\ee
Without loss of generality, let us first consider the special
case of $\xi^a = \xi \d^{a3}$. To diagonalize $\cK$, we note that
\be
M := J_3 + S_3
\ee
is  conserved. We can diagonalize $\cK$ in sectors of fixed $M$.
It is convenient to work with the eigenvectors of $S_3$: 
\be \label{e0pm}
  e_a^{(0)}=n_a,
  \qquad
  e_a^{(\pm)}=\frac{e_a^{(1)}\pm i e_a^{(2)}}{\sqrt2}, \qquad n_a = \d_{a3}
  \ee
\be
  S_3e^{(0)}=0,
  \qquad
  S_3e^{(\pm)}=\pm e^{(\pm)}.
\ee
At fixed $M$, the most general vector link is thus given by
\be
W_a^{(M)}=\left( w_M^0 e_a^{(0)}|J,M\rangle
+w_M^+ e_a^{(+)}|J,M-1\rangle+w_M^- e_a^{(-)}|J,M+1\rangle \right)
\bra{\xi, \hat{z}}
\label{fixedMstate}
\ee
where $w_m^{0}, w_M^{\pm}$ are arbitrary coefficients.
Here it is understood that a
component is absent if the required orbital magnetic number $m$ of
$J_3$ lies outside $-J\leq m\leq J$.
In the ordered basis $(w_M^0,w_M^+,w_M^-)^T$, after a straightforward
computation, we obtain
\be
\cH_M(\xi)=2
\begin{pmatrix}
J(J+1)-M^2 & (\xi-M+1)A_M & (\xi-M-1)B_M\\
(\xi-M+1)A_M & (\xi-M+1)^2+A_M^2 & -A_MB_M\\
(\xi-M-1)B_M & -A_MB_M & (\xi-M-1)^2+B_M^2
\end{pmatrix},
\label{Hmatrix}
\ee
and
\be
(\vect J\cdot\vect S)_M=
\begin{pmatrix}
0&-A_M&B_M\\
-A_M&M-1&0\\
B_M&0&-(M+1)
\end{pmatrix}.
\label{JSmatrix}
\ee
where
\be
  A_M :=\sqrt{\frac{(J+M)(J-M+1)}{2}},
  \qquad
  B_M :=\sqrt{\frac{(J-M)(J+M+1)}{2}}.
\ee
Noticing that $A_M, B_M=0$ when $M=J+1$ or $-(J+1)$, we  observe that 
the diagonzlization problem depends on the values of $M$ as described
in the table \ref{ttt}
\begin{center} \label{ttt}
\begin{tabular}{@{}lll@{}}
\toprule
Sector & Present components & Problem\\
\midrule
$-J+1\leq M\leq J-1$ & $0,+,-$ & $3\times3$ diagonalization\\
$M=\pm J$ & two components & $2\times2$ diagonalization\\
$M=\pm(J+1)$ & one component & exact $1\times1$ eigenmode\\
\bottomrule
\end{tabular}
\end{center}

We note that for
\be
  M=J+1,
\ee
only the positive helicity component is allowed in \eq{fixedMstate}
and we obtain an exact eigenmode
\be \label{outmode}
  W_a
  =w e_a^{(+)}|J,J\rangle_H{}_E\langle\xi,\hat z|
  \ee
with the  eigenvalue 
\be
  k_-(\xi)=2(\xi-J)^2-4(J+1).
\label{kminus}
\ee
This {\it outward extremal mode}  is tangential since
$n^ae_a^{(+)}=0$. 
Using $\Tr W_a^\dagger W_a=|w|^2$, 
we finally obtain for the  quadratic potential 
\be \label{V2-xi3}
  U_2(\xi)= \a |w|^2 k_-(\xi).
\ee
As for the quartic part, we obtain from \eq{outmode} that
\be
  q^H_{ab}=|w|^2 D_{ab}|J,J\rangle_H {}_H\langle J,J|,
  \qquad \mbox{where}\quad
  D_{ab} :=e_a^{(+)}e_b^{(+)*}-e_b^{(+)}e_a^{(+)*}.
\ee
In general, the circular polarization with respect to a general
normal vector $n_a$ satisfies
\be \label{ee-P}
e_a^{(+)}e_b^{(+)*} = \frac{1}{2}(P^T_{ab} + i \e_{abc} n_c),\qquad
P^T_{ab} := \d_{ab} -n_a n_b.
\ee
This means $D_{ab} = -i \e_{abc}n_c$,
$\sum_{a,b} D_{ab} D_{ab} = -2$ and so 
\be \label{V4}
U_4 = 2 \a |w|^4.
\ee
Therefore we obtain the potential
\be \label{Uw}
U(w) =  \a (k_-(\xi) |w|^2 + 2 |w|^4)
\ee
and it is interesting that  there
is a nonvanishing condensate
\be
|w|^2 = -\frac{k_-(\xi)}{4}
\ee
whenever the eigenvalue $k_-$ is negative. The configuration
\be
\xi_a =J n_a
\ee
with the lowest energy is preferred.
The corresponding condensate and potential $U$ is given
by
\be
|w|^2 = J+1, \qquad U = -\frac{(J+1)^2}{l_P^2}.
\ee

So far this is for a single mode at the north pole. Using
rotation, we can immediately 
get a mode at the ``rotated  north pole'' $\O \in S^2$: 
\be \label{Z-rot}
W_a(\xi, \O) = w(\O) e_a^{(+)} \ket{\O}_E {}_H\bra{\xi, \O}.
\ee
We can then get a rotational invariant configuration by uniformly
populating the spherical surface with \eq{Z-rot} and the total potential
is obtained by summing over the contributions of each individual mode.
For the rotated mode \eq{Z-rot}, the potential has the same form
\eq{Uw}
due to rotational
covariance of the quadratic operator \eq{cK}.
The corresponding condensate vacuum prefers the configuration
\be \label{lowest-xi}
 \xi_a = J n_a.
 \ee
 Substituting $\xi_a$, \eq{xi-L-y} reads
 \be \label{rj}
\frac{r}{2l_P} = \sqrt{(J-Lp_r)^2 +(L p_\parallel)^2}.
 \ee
 Let us consider a rotational invariant static condensate and so
 \be
 p_\parallel =0, \qquad p_r =0.
 \ee
 However one should be careful with the vanishing condition for $p_r$.
 Note  that as the radial direction has been effectively regularized to
 have a size $L$,  a radial momentum satisfying
 $|p_a| \lessapprox  1/L $ should be considered effectively
 a vanishing momentum.
Now the links are situated outside
 the horizon and so  $r = |x_a| \geq R= N l_P$. Therefore
the condition \eq{rj} implies that
\be
r = R  + \delta, \quad \delta \lesssim l_P.
\ee
As a result, we obtain a
condensate of links uniformly populating a  thin region of width $l_P$
located right outside the horizon, whose amplitude is
$|w|^2 = \frac{N}{2}$
in the leading order of large $N$.

Next we turn on the gauge fields. We can write
\be
h_{ab} = \cD_a W_b - \cD_b W_a = \bar{h}_{ab} + \d h_{ab} 
\ee
with
\be
\bar{h}_{ab} :=w(\xi_a \Psi_b - \xi_b \Psi_a), \quad
\d{h}_{ab} := w (C_a \Psi_b - C_b \Psi_a),
\ee
where we have denoted $\Psi_b (\O) := e_b^{(+)} \ket{\O}\bra{\xi,\O}$
for each mode at the position $\O$ and
\be
C_a := R(B_a - A_a) 
\ee
is the difference of the gauge fields.
We are interested in the the gauge-field-dependent potential
in \eq{V2} arising from $\d{h}_{ab}$. 
Using \eq{ee-P}, it is easy to  obtain
$\sum_{a,b} \d h^\dag_{ab} \d h_{ab} = |w|^2 (C_A C^A + 2 C_\perp^2)$,
$\sum_{a,b} \bar{h}^\dag_{ab} \d h_{ab} = 2  |w|^2 (J-\xi_\perp)
C_\perp =0$.
Therefore  we obtain the interaction potential from $U_2$
\be
U_{\rm int, mode}(C)= \a |w|^2 \Big[
  C_A C^A + 2 C_\perp^2
%c5 + 4 (J-\xi) C_\perp
  \Big]
\ee
from each mode at $\O$. Here we have used the condition \eq{lowest-xi}.
Summing over the cells over the fuzzy sphere using \eq{TrN},
we obtain finally
\be \label{VC}
V_{\rm int} (C) = \frac{\cK}{2}\int d\O \left[
  (b_A - a_A )^2 +
  2 (\upvarphi_H - \upvarphi_E )^2
\right]
\ee
where
\be
\cK :=  \frac{N^3 l_P}{8\pi a_0}
\ee
is a stiffness
constant that governs the depth of the potential.
Since $\cK \sim O(N^3)$ is
large, the cost is high to fluctuate over the minimal configuration
\be \label{locking-ba}
b_A = a_A , \qquad
\upvarphi_H = \upvarphi_E.
\ee
However kinetic motion can move the fields away from the minimum.
We will next show that the potential term is indeed much larger compared
to the kinetic terms in our model and so the gauge fields are locked.

\subsection{Quantum membrane  Ohmic, Hall and polarization current }

Let us consider the full black hole-environment system.
As we are primarily
interested in the EM boundary
condition, let us drop the scalar parts for this discussion.
The action reads
\be
S = -\frac{1}{16\pi} \int d^4 x \sqrt{-g} F^2
+ \frac{R^2}{2 g_{\rm BH}^2} \int d\t d\O 
\left(  (\del_\t b_A)^2 -\frac{1}{l_P^2} f^2\right)
%c13 - \int d\t d\S\; j_A b_A
+S_{\rm parton}[\psi,b]
-  \frac{\cK}{2} \int d\t d\Omega\; (b_A -a_A )^2 ,\;\;
\ee
where the condensation of the $W$-links has been incorporated.
The locking mass term \eq{VC} gives a strong coupling of the gauge field
on the fuzzy sphere with the gauge field of the environment. 
Variation
with respect to $a_A$ and $b_A$ gives the equation of motion respectively
\bea 
&& \frac{1}{4\pi} F_B{}^n + \frac{\cK}{R^2} (b_B- a_B) =0 ,
\label{ll1}\\
&& \frac{1}{g_{\rm BH}^2} (\Box b_B -\del_B \del_A b_A) 
-j_B - \frac{\cK}{R^2} (b_B- a_B) =0, \label{ll2}
\eea
where $\Box = \del_A^2 - \del_\t^2$, $\del_A := l_P^{-1} D_A$ and
\be
j_A := -\frac{1}{\sqrt{-h}}\frac{\d S_{\rm parton}}{\d b^A}
\ee
is the parton current \eq{jA}.
Here the  superscript is included to indicate that
the electric field is determined by the gauge potential $b_A$ of the
fuzzy sphere:
\be
E^{(b)}_A = -\del_\t b_A.
\ee
The equations \eq{ll1} and \eq{ll2} give immediately
\be \label{ll3}
\frac{1}{4\pi} F_B{}^{n} = j_B^{\rm mem},
\ee
where
\be \label{c1}
 j^{\rm mem}_{A} = j_{A} + j^{\rm pol}_{A}, \qquad
 j_A^{\rm pol} :=  -\frac{1}{g_{\rm BH}^2}
 \left(\Box \d_{AB}-\partial_A\partial_B\right)b^B. 
 \ee
$j_A^{\rm pol}$ is a polarization current that arises from the kinetic motion
  of the gauge field.
\eq{ll3} is the matrix model version of the boundary condition \eq{js}.
%c13
In the FIDO frequency domain
$b_A \sim e^{-i \oh \t}$, \eq{ll2} can be solved
to give a differential equation 
\be
\left((\frac{\cK}{R^2} -\frac{1}{g_{\rm BH}^2}(\oh^2 -p^2)) \d_{AB}
-\frac{1}{g_{\rm BH}^2}p_A p_B + i\oh 
(\s_{xx} \d_{AB} + \s_H \e_{AB})\right)b^B
= \frac{\cK}{R^2} a_A.
\ee
Now $\cK/R^2$ is of order $N$, $1/g_{\rm BH}^2$ is of order 1
and $\s_H$ is at most of order 1, 
therefore as long as the
fields are slow varying with
\be \label{ccc1}
\hat \omega \ll \frac{\sqrt{N}}{l_P}, \qquad |p_A| \ll \frac{\sqrt{N}}{l_P},
\ee
one can ignore the kinetic terms in the kernel and obtain
\be \label{locked-ba}
b_A = a_A (R^+).
\ee
Note that the
%c13
coordinate thickness of the stretched horizon $ \d r \sim l_P$
is much smaller than the radial extent $L= N' l_P/N$ of the
environmental Maxwell theory. Thus  \eq{Maxwell} provides
a good description of the physics outside the horizon that well covers
the stretched  horizon and we can trust our derivation of condensate.
%c13
As a result, \eq{ll3} gives
\be \label{ll4}
F_B{}^n = 4\pi j_B^{\rm mem}, \qquad j_B^{\rm mem} = \s_{BC}(\oh) E^C,
\ee
where the current is now a direct response of the bulk electric field,
with $\s_{BC}$ being the electrical admittance of the horizon:
\be
\s^{BC}(\oh) :=  ( \s_{xx} + i\oh \Dh ) \d_{AB}
+ \s_H \e_{AB} ,
\ee
where
\be \s_{xx} = \frac{1}{4\pi}, \quad
\Dh := 
\frac{l_P}{2\pi a_0}, \quad \s_H = \frac{q Q}{2\pi N}.
\ee
Here the Ohmic term $\s_{xx}$ originates from the dissipative effects
of the horizon, while the Hall term $\s_H$
arises from the LLL transport of the
partons induced by the intrinsic Berry monopole on the fuzzy sphere.
The $\Dh$ term admits a  natural interpretation as a polarization current.
Indeed, when the external electric field varies in time,
the locked horizon field cannot respond instantaneously. The kinetic
term of the horizon gauge field then produces a reactive current that
opposes the change of the electric field. This current may be written as
a polarization current
\be
j_A = \del_\t P_A,
\ee
with a polarization density
\be
P_A =  \chi E_A, \qquad \chi = -\Dh <0.
\ee
The fact that $P_A$  points opposite to $E_A$ suggests that
the stretched horizon  behaves not as an ordinary dielectric composed of
polarizable charges, but as a reactive quantum surface with
negative electric susceptibility.
Both
the Hall current and the polarization current are nondissipative.
Our result \eq{ll4} generalizes the classical membrane Ohm's law \eq{EAFA}.

%c13
The black hole boundary condition is usually expressed in terms of
the Schwarzschild time $t$. Using $t$, the FIDO frequency $\oh$ becomes  the
asymptotic frequency 
\be
 \omega :=\frac{2a_0 M_P l_P}{N} \hat\omega.
 \label{frequency_redshift}
 \ee
The locking condition \eq{ccc1} may  be written as
\be
 \omega\ll\frac{1}{\sqrt{N} l_P },
 \qquad
 L\ll\sqrt N,
 \label{locking_asymptotic}
 \ee
 where we have used $p_A^2 = L(L+1)$ for an angular momentum mode $L$. And the
 membrane current now read
 \be
 j_B^{\rm mem} = \s_{BC}(\o) E^C,
 \ee
where
\be \label{cc1}
\sigma_{AB} =(\sigma_{xx}+ i \o D)\delta_{AB} +\sigma_H\epsilon_{AB}, \qquad D = 
\frac{N}{4\pi a_0^2 M_P}, \quad \s_H = \frac{q Q}{2\pi N}. 
\ee
This fixed-parton transport is valid at low energies
\be
\o R \ll 1.
\ee

Summarizing, as a result of the  condensation of the off-diagonal
link matrices, the horizon LLL degrees of freedom respond directly
to the boundary value of the bulk Maxwell field,
thereby  making the electrical
current of the black hole horizon a tangible  quantity measurable to
the outside observer. Thus our model provides a microscopic
derivation of the stretched horizon, turning the fictitious
membrane in the original membrane paradigm dynamical and physical,
opening up opportunities to probe quantum gravity through the
structure of the quantum horizon. We will investigate one such
possibility next. 

\subsection{Reflective boundary condition for the quantum horizon}

Although we have not derived the spacetime Schwarzschild metric from our
model, this must happen if we assume general covariance does emergence
in our model. We leave this important question for further investigation.
For now, let us assume that we do have a Schwarzschild metric and ask the
question of how a quantum horizon as described by our model, 
particularly the generalized Ohm's law \eq{ll4}, will
modify the boundary condition for EM modes.

As before, we express the electric field in terms of the Schwarzschild
time $t$ using \eq{EFf}. And using \eq{Fnr} for
the normal component of the field strength, we arrive finally at
\be \label{GBC}
\del_{r_*} a_A -4 \pi \s_{AB} \del_t a_B =0
\ee
It is convenient to diagonalize the boundary condition \eq{GBC},
or equivalently the generalized Ohm's law \eq{ll4}. To do this, 
let us use a local oriented orthonormal frame  $e_A^{1},e_A^{2}$
on the horizon so that $\e_{12} =+1$. 
Introduce the right- and left-helicity polarization vectors
\be
    e_A^{(+)}:=\frac{1}{\sqrt{2}}\left(e_A^{1}+\ii e_A^{2}\right),
    \qquad
    e_A^{(-)}:=\frac{1}{\sqrt{2}}\left(e_A^{1}-\ii e_A^{2}\right).
    \label{helicitybasis}
\ee
They obey
\be
    \e_A{}^B e_B^{(\pm )} =  \pm \ii e_A^{(\pm)}.
    \ee
As a result, 
we obtain the diagonalized membrane Ohm law in the helicity basis:
\be
j^{(\pm)}_A = \s_\pm E^{(\pm)}_A, \qquad \s_\pm := \s_{xx} + i(\o D \pm \s_H)
\ee
for a right or left polarized electric field $E^{(\pm)}_A$.
Note that the Hall term splits the two helicities with opposite signs.
As a result, the boundary condition is diagonalized 
\be \label{GBC1}
\del_{r_*} a_A^{(\pm)} -4\pi \s_\pm \del_t a_A^{(\pm)} =0.
\ee
This can be rewritten as a relation between the ingoing and the
outgoing wave:
\be
\del_u a_A^{(\pm)} = \cR_\pm \del_v  a_A^{(\pm)} ,
\ee
where
\be
u = t-r_*, \quad v = t+r_*
\ee
are the outgoing and ingoing null coordinates for
the Schwarzschild metric, and 
\be
\cR_\pm = \frac{-i D_\pm}{1+i D_\pm}, \qquad
D_\pm :=  2\pi (\o D\pm \s_H)
\ee
is the reflection coefficient.
In the case $D= \s_H =0$, we obtain  $\del_u a_A^{(\pm)} =0$ and
perfectly absorbing horizon is recovered.
It is interesting that the quantum
horizon is no longer perfectly absorbing, but has a reflectivity
controlled by the polarization and the Hall coefficient of the
horizon. 
In general the reflection is frequency dependent 
and demonstrates a chiral response for a charged black hole where
$\s_H \neq 0$.

For general $\o$ and $Q <N$, we can write $\cR_\pm$ in terms
of an amplitude and a phase factor
\be
\cR_\pm = \frac{D_\pm}{\sqrt{1+ D_\pm^2}} e^{-i (\frac{\pi}{2}+\d_\pm)},
 \qquad \tan \d_\pm = D_\pm.
 \ee
 The difference between $\d_+$ and $\d_-$ implies a 
horizon induced Kerr rotation where a
 linearly polarized incident wave is reflected as a rotated
 and elliptically polarized wave.
The rotation angle of a reflected linear polarization is
 \be
 \th_{\rm rot} = \frac{1}{2}(\d_- -\d_+) \approx -2\pi \s_H,
 \ee
where the approximated value is for small $\s_H$, $\o D$.
And the weak ellipticity angle is
\be
\eta_{\rm ellp} = \frac{1}{2}\log \left|\frac{\cR_+}{\cR_-}\right|
\approx \frac{\s_H}{\o D},
\ee
where the last expression is expanded around small $Q$.

We remark that a partial reflection at the stretched horizon
naturally leads to black-hole echoes:
electromagnetic waves reflected by the quantum
horizon can undergo repeated scattering between the near-horizon
surface and the exterior photon-sphere barrier, producing a sequence
of delayed signals. Such echoes are often studied
phenomenologically \cite{Berti:2025hly,
Cardoso:2016rao,Cardoso:2016ryw,Abedi:2016hgu,
Cardoso:2017cqb,
Bambi:2015kza,Conklin:2017lwb,
Wang:2019rcf
}
by postulating an effective reflection coefficient.
As a result, the prediction carries a model uncertainty which
limits the extent to which a deviation in the data can be unambiguously
interpreted as evidence for new physics \cite{Ashton:2016xff,
Westerweck:2017hus}.
On the contrary, in our microscopic horizon model, the
reflection coefficient is not introduced as an arbitrary
phenomenological function, but is derived from the membrane current
response. The echo signal is therefore tied to the transport
coefficients of the quantum horizon, reducing the freedom of the fit.
It may be interesting to redo some of the echo phenomenology
analysis in the light of our model of quantum horizon as a
possible more direct test of the
underlying quantum-gravity dynamics.

\section{Discussions}

In this paper, we have shown that the $N$ flavor of fermions in our model
with a multiplicative Yukawa coupling resulted in a system of LLL partons.
The LLL partons not only contribute to the
energy and entropy counting of the black hole, but also give rise to
novel current on the horizon. We have also derived a novel
locking of gauge fields from the dynamics of the
matrix model off-diagonal links.
Combing these results, we derived microscopically
membrane paradigm for quantum black hole with
a membrane/stretched horizon occupying a thin
region of space where the link fields condense, and the membrane partons
obey a first order LLL dynamics govern by the intrinsic Berry monopole
on the fuzzy sphere.

The LLL membrane current receives in addition to an Ohmic component, but
also a Hall component and a polarization component.  With our model,
the membrane current is real with clear physical origin for its
electrical properties.  The Ohmic resistivity is originated from the
dissipative horizon heat bath where the partons live. The Hall
conductivity arises from the presence of a Berry monopole that is
intrinsic on the fuzzy sphere and that is visible to the LLL
partons. The polarization current arises from the kinetic term of the
worldvolume gauge field.  As a result of the parton dynamics, the
horizon become a quantum noncommutative manifold with a frequency and
charge dependent reflection coefficient. 

%c13
This quantum-membrane  description was derived in the low energies regime
$\omega R \ll 1$ in which the total parton number remains fixed. At
frequencies of order the parton-production threshold, $\omega R =O(1)$,
real parton excitations become possible and additional scattering channels open:
\be
\gamma(\o)+{\rm BH}_i\longrightarrow
\begin{cases}
\gamma_{\rm ref}(\o)+{\rm BH}_i,
& \text{elastic reflection},\\[1mm]
{\rm BH}_f+X_f,
& \text{inelastic absorption}.
\end{cases}
\label{S_channels}
\ee
Consequently, the fixed-parton-number constitutive description is no longer
complete. The appropriate microscopic object is then the matrix-model
black hole S-matrix, which encodes the full set
of elastic and inelastic scattering amplitudes. Universal constraints on this 
S-matrix may therefore provide a powerful framework for probing
the deeper quantum physics of black holes.

The locking mechanism has a natural open-string interpretation. In the
two-block matrix model, the off-diagonal blocks transform as
bi-fundamentals under the black-hole and exterior gauge groups, and
may therefore be viewed as open strings stretching between the
fuzzy-sphere horizon and the exterior sector. When these modes become
tachyonic near the stretched horizon, their condensate acts like an
open-string tachyon condensate. Since the condensate is charged under
both sectors, it gives a Higgs mass to the relative gauge field
$b_A-a_A$, thereby enforcing the locking of gauge fields.  This
suggests that our quantum model of black hole may have a
string-theoretic origin, see \cite{Laurenzano:2025vfh} for a
suggestion.  It would be interesting to look for a construction of
black hole in terms of branes and obtain the black hole model
\cite{Chu:2024qil} from string theory.

A possible route toward gravity is to regard geometry as
an emergent, long-wavelength description of matrix dynamics.
Around a matrix background $X^{(0)}_a$, the commutator with a
low-energy probe defines an effective differential operator
\be
- i [X_a, \Phi] \longrightarrow e_a{}^\nu(x) \del_\nu \Phi + \cdots,\qquad
a=0,\cdots , 3
\ee
so that the coefficient of the derivative may be interpreted as an
emergent vielbein, while the principal symbol of the matrix Laplacian
determines an effective inverse metric,
\be
-[X_a,[X_a,\Phi] \longrightarrow -g^{\m\n}(x) \del_\m \del_\n \Phi +\cdots,
\qquad g^{\m\n} = \sum_a e_a{}^\m e_a{}^\n.  
\ee
The collective modes that change the
kinetic operator are candidates for vielbein or metric fluctuations.
It is very likely that the 1-loop induced action from
the matrix model dynamics will contain the
Hilbert-Einstein term as its leading two-derivative contribution.
However one should be careful about ghost \cite{Ho:2025htr}.
It will also be interesting to derive the ``close-string''
GR membrane paradigm
by considering the stress tensor of the
microscopic degrees of freedom of the theory
and transfer the geometric response of the horizon
to become boundary condition for gravity by the off-diagonal block link
condensate.

Let us comment on the possible relation between
our model and the fuzzball proposal \cite{Mathur:2005zp}.
Both approaches suggest that the classical horizon may be an effective
description of a very large number of microscopic degrees of freedom
rather than a fundamental property of an exact quantum state.
In the fuzzball picture, black-hole microstates are conjectured to be
horizon-scale and horizonless, although only restricted classes
admit smooth supergravity descriptions. 
Since fuzzball has no conventional vacuum horizon,
its radiation does not arise microscopically through the usual
semiclassical Hawking pair-creation process.
Nevertheless, its
response can reproduce approximately thermal emission.
Apparent absorption and an approximately ingoing boundary condition
may instead arise at finite temporal or spectral resolution,
when an incident perturbation excites a large number of
internal states whose
subsequent emissions are delayed and dephased \cite{Martinec:2023iaf}.
\footnote{We thank Emil Martinec for a discussion of this point.}
Several analogous features find  a concrete realization in our model.
The microstates are many-body LLL states on the fuzzy sphere, with
coherent wave packets resolving areas of order $\Delta A\sim
l_P^2$. Hawking-like decay arises not from pair creation across a
smooth horizon, but from rank-changing quantum tunneling accompanied
by monopole creation. This mechanism is distinct from, though possibly
complementary to, the semiclassical spacetime tunneling picture of
Ref.~\cite{Parikh:1999mf}. Likewise, the interacting LLL partons
provide a microscopic origin for absorption and the effective ingoing
boundary condition. 
It is therefore interesting to ask whether our model may be viewed
as a possible microscopic realization of the fuzzball paradigm, or
whether the two represent distinct mechanisms that reproduce the same
coarse-grained horizon physics.

In general relativity, black hole is obtained as a vacuum solution of
the Einstein equation, with the same vacuum metric inside and out.
From the point of view of our matrix model, it is natural
to consider the fuzzy sphere as a domain wall (a crack for quantum space).
Then it is more natural that we can have a  different vacuum solution
in the interior of the horizon. In our matrix model, this can be included
with a third interior block $X_a^{\mathrm I}$. It will be interesting
to analysis this 3-blocks matrix model using the multi-block
model developed here. In  this case, new
interaction vertices arise that are missing in 
the 2-bodies interaction. It will be interesting to see if the interior
block has a geometric description and what.
It will also be extremely interesting to formulate and resolve
the puzzles of black hole like Page curve or singularity from the
matrix model.

%c12
%% Finally let us  comment on the important roles play by the
%% negative quadratic term in \eq{L} of the model.
%% While around
%% the trivial commuting background it is tachyonic, this is precisely
%% what drives the system toward a nontrivial fuzzy sphere background.
%% Around the
%% fuzzy-sphere background, its effect is reorganized by the
%% non-abelian interactions and by quantum corrections. A full classical
%% and quantum analysis
%% show that the fuzzy sphere is perturbatively stable \cite{chu-qs}. This
%% leaves behind a metastable effective potential for the monopole
%% induced tunneling process. The same tachyonic structure also
%% supplies the soft dissipative effects needed to motivate an effective
%% temperature and relaxation rate in the derivation of the Ohmic
%% membrane current. Moreover, when the black-hole block is coupled to an
%% environmental block, the negative mass term produces tachyonic
%% off-diagonal link modes localized near the horizon. Their condensation
%% locks the external gauge field to the horizon gauge field, thereby
%% turning the microscopic response of the fuzzy-sphere partons into the
%% macroscopic membrane boundary condition. Thus the negative mass term
%% is responsible simultaneously for tunneling, dissipation, link
%% condensation and the emergence of the membrane paradigm.

%c12
In our model, we suspect that black hole quantum
chaos \cite{Susskind:2014rva,
Shenker:2013pqa,
Maldacena:2015waa,Maldacena:2016hyu} may admit a natural realization
through the many body dynamics of the horizon partons.
As shown in this work, the horizon partons in our model
form an LLL system with quantum-Hall kinematics. Upon including
interactions, the resulting theory may resemble an interacting
fractional quantum Hall (FQH) system. Since the kinetic energy is
quenched within the highly degenerate lowest Landau level, the
dynamics is governed primarily by the projected interactions, and
sufficiently generic interactions are expected to produce
nonintegrable many-body evolution. This expectation is supported by
exact-diagonalization studies of LLL-projected fractional quantum Hall
Hamiltonians, in which Wigner--Dyson level statistics, characteristic
of quantum-chaotic spectra, has been observed \cite{Fremling_2018}.
In this regard, an important problem is therefore to identify the
interaction channels in the matrix model that efficiently mix the
degenerate LLL states and scramble quantum information over the fuzzy
horizon. Establishing nonintegrability through spectral statistics,
and independently diagnosing scrambling through out-of-time-ordered
correlation functions, would provide complementary tests of horizon
quantum chaos.

%c9
%% From our discussion, we are led to conjecture that
%% it is a fundamental property of
%% quantum spacetime that the conductivity for horizon
%% \be
%% \s_{xx} = \frac{n}{4\pi}, \quad n =  \pm 1, \cdots
%% \ee
%% is quantized in quantum gravity, with $n=1$ for the black hole horizon
%% and $n=-1$ for the cosmological horizon. In fact

A classical
GR membrane paradigm for the cosomological horizon has been discussed in
\cite{Wang:2014aty}
and it was found that a similar fluid description exists. 
For the EM case, let us derive the
corresponding classical membrane paradigm. 
Let us consider
the de-Sitter static patch
\be
ds^2 = -f(r) dt^2 + f(r)^{-1}dr^2 +r^2 d\O^2,\qquad f(r) = 1 -r^2/L^2
\ee
and put a cosmological stretched horizon at $r=H^{-1} -\e$, a static FIDO
has $u^\m = f^{-1/2} \del_t^\m$, $n^\m = f^{1/2} \del_r^\m$ where $n^\m$
points outward, toward the unobservable region beyond the cosmological
horizon.
Then it is easy to see that the {\it outgoing condition}
from the static patch ($\del_t + \del_{r_*}) A_a =0$ where the tortoise
coordinate is $r_* = \frac{L}{2} \ln(\frac{L+r}{L-r})$)
is equivalent to the membrane current
\be
j_A^{\rm cosmo} = -\frac{1}{4\pi} E_A
\ee
where $E_A$ is the FIDO  electric field $E_A = F_{A \hat{t}}$.
Here the minus sign is a reflection of the orientation:
for a black-hole horizon, energy flows inward into the hole;
for a cosmological horizon, energy flows outward,
leaving the observer’s static patch. This result suggests that there is a
quantum membrane description of the cosmological horizon
in our matrix model. Working out the quantum currents is interesting and
will give ways to show quantum gravity effects of the cosmological
horizon, which may leave imprints through the
cosmological collider \cite{Arkani-Hamed:2015bza}.

\appendix

\section{Continuum and Fuzzy Two-Sphere}
In this appendix, we spell out the notations and conventions
for the geometry of the continuum and fuzzy two-sphere we use
in the paper.

\noindent \underline{Continuum sphere}

Let $S_R^2\subset\mathbb R^3$ be described by
\be
 x^a(\xi)=R\,n^a(\xi),
 \qquad
 n^a n_a=1,
 \qquad
 \xi^A=(\theta,\phi),
\ee
where $a=1,2,3$ is an ambient Cartesian index and $A=\theta,\phi$ is a
coordinate index on the sphere.  In standard spherical coordinates,
$ n^a=(\sin\theta\cos\phi,\,\sin\theta\sin\phi,\,\cos\theta)$.
The metric
on the sphere of radius $R$ is
\be
 g_{AB}=R^2 h_{AB},
\ee
where
\be
 h_{AB}:=\partial_A n^a\,\partial_B n_a
\ee
is the unit-sphere metric.

The dimensionless embedding tangent frame is
\be
 e_A{}^a:=\partial_A n^a,
 \qquad
 e_a{}^A:=h^{AB}e_{Ba}.
\ee
It obeys
\be
 n_a e_A{}^a=0,
 \qquad
 e_A{}^a e_{Ba}=h_{AB},
 \qquad
 e_a{}^A e_A{}^b=P_a{}^b,
\ee
where
\be
 P_a{}^b:=\delta_a{}^b-n_a n^b
\ee
is the projector onto the tangent plane.
The physical embedding vielbein is
\be
 E_A{}^a:=\partial_A x^a=R e_A{}^a,
 \qquad
 E_a{}^A=\frac1R e_a{}^A,
\ee
so that
\be
 g_{AB}=E_A{}^aE_{Ba}.
\ee
Another useful frame is the Killing tangent frame  defined by
\be
 K_A{}^a:=\epsilon^{abc}n_b e_{Ac}
 =\epsilon^{abc}n_b\partial_A n_c.
\ee
It is tangent,
\be
 n_aK_A{}^a=0,
\ee
and is the $90^\circ$ rotation of the embedding tangent frame:
\be
 K_A{}^a=\varepsilon_A{}^B e_B{}^a,
\ee
where $\varepsilon_A{}^B$ is the complex structure on the oriented two-sphere.
The Killing frame satisfies
\be
K_A{}^aK_{Ba}=h_{AB},
 \qquad
 K_a{}^AK_A{}^b=P_a{}^b,
 \qquad
 \nabla_A K^A_a =0,
\ee
where $K_a{}^A:=h^{AB}K_{Ba}$ and $\nabla_A$ is the covariant
derivative on the sphere. 

The three rotational vector fields are given by
\be
 \cK_a:=K_a{}^A D_A,
\ee
and the corresponding Hermitian angular-momentum operators may be chosen as
\be
 L_a:=-iK_a{}^A\nabla_A.
\ee
and obey the algebra
\be
 [L_a,L_b]=i\epsilon_{abc}L_c.
\ee
%% The covariant derivative $\nabla_A$ is defined as usual. For example, 
%% \be
%%  \nabla_A f=\partial_A f,
%% \ee
%% for a scalar $f$,
%% and $\nabla_A V^B=\partial_A V^B+h^B{}_{AC}V^C$ for a vector $V^B$.
%% %c  with the help of the Christoffel symbol. 
  
Any ambient three-vector field $B_a(\xi)$ can be decomposed using the
embedding frame
as
\be
 B_a=e_a{}^A \tilde{b}_A+n_a\upvarphi, \quad
\mbox{with $\tilde{b}_A:=e_A{}^aB_a$ and $\upvarphi:=n^aB_a$},
\ee
or equivalently using the Killing frame,
\be
 B_a=K_a{}^A b_A+n_a\upvarphi.
\ee
The two tangent one-forms are related by a Hodge rotation,
\be
 b_A=-\varepsilon_A{}^B \tilde{b}_B,
\ee
up to the chosen orientation convention.

\noindent \underline{Fuzzy sphere}

For finite $N$, the fuzzy sphere $S_N^2$ is defined as a 
spin-$J$ irreducible representation $J_a$ of $SU(2)$,
\be
 [J_a,J_b]=i\epsilon_{abc}J_c,
 \qquad
 J_aJ_a=J(J+1)\id_N,
 \qquad
 N=2J+1.
\ee
A function on the fuzzy sphere is an $N\times N$ matrix $\widehat f$.
The rotational derivative is defined as, for example,
\be
  \cL_a  F:=[J_a,  F], \qquad
  \cL_a  \psi = J_a \psi,
\ee
for  function $F =(F_{mn})$ and vector
$\psi = (\psi_m)$.
These operators obey
\be
 [\cL_a,\cL_b]
 =i\epsilon_{abc} \cL_c.
\ee
In the large-$N$ symbol map,
\be
 F \longleftrightarrow F(\theta,\phi),
 \qquad
 [J_a, F]\longrightarrow L_a F.
\ee
The fuzzy Laplacian is
\be
 \Delta_N  F
 :=\frac1{R^2}[J_a,[J_a, F]].
\ee
It approaches the positive Laplace operator $-\nabla^2$:
\be
 \Delta_N \longrightarrow-\nabla^2.
\ee
The fuzzy spherical harmonics $Y_{l m}$ satisfy
\be
 [J_a,[J_a, Y_{l m}]]
 =l(l+1)  Y_{l m},
 \qquad
 0\leq l \leq 2J=N-1.
\ee
Hence finite matrix size gives an angular-momentum cutoff at $l_{\max}=N-1$.
At large $N$, trace becomes the integral over $S^2$:
\be
 \frac1N\Tr F
 \longrightarrow
 \frac1{4\pi}\int_{S^2}\dd\Omega \; F,
 \qquad
 \dd\Omega=\sin\theta\,\dd\theta\,\dd\phi.
\ee
Equivalently,
\be
 \Tr F
 \longrightarrow
 \frac{N}{4\pi R^2}
 \int_{S_R^2}\dd^2\xi \,\sqrt g\,F.
\ee

A fuzzy vector fluctuation is a triplet of matrices $ B_a$.
In the continuum limit, it decomposes as
\be
  B_a
 \longrightarrow
 e_a{}^A \tilde{b}_A+n_a\upvarphi
\ee
or equivalently
\be
 B_a
 \longrightarrow
 K_a{}^A b_A+n_a\upvarphi.
\ee
Thus the large-$N$ limit of a three-component matrix fluctuation consists
of a two-dimensional tangent gauge field and a normal scalar.

\section{Coherent State on Fuzzy Sphere}

Consider the fuzzy sphere defined by a spin $J$ representation of the
$SU(2)$ algebra
\be \label{JJ}
[J_a, J_b] = i \e_{abc} J_c, \qquad N= 2J+1.
\ee
To facilitate a
precise bridge between the finite $N$ matrix degrees of freedom and
the ordinary fields on a classical sphere in the large $N$ limit,
we introduce the $SU(2)$ coherent states
defined by
\be
\ket{\Omega} := g(\Omega) \ket{J,J},
\ee
where $\Omega = (\th, \phi)$ represents a point on $S^2$,
$g(\Omega) \in SU(2)$ is an rotation taking the north pole to $\Omega$,
$J_a$ is the angular momentum algebra \eq{JJ}, 
and $\ket{J,m}, m = -J, -J+1, \cdots, J$ are angular momentum basis
states. It follows that
\be
\bra{\Omega} J_a \ket{\Omega} = J n_a (\Omega),
\ee
where $n_a = (\sin \th \cos \phi, \sin \th \sin \phi, \cos \th)$
is the outer normal to the sphere at the point $\Omega$.  
In the leading order of large  $N$, the fuzzy sphere coordinates satisfies
\be
\bra{\Omega} J_a \ket{\Omega} \simeq R n_a,
\ee
where $R= N/2$ is the dimensionless radius of the fuzzy sphere. Thus the
coherent state provides a matrix analogue of a state localized near the
point $\Omega$ on the sphere and
thereby allows matrix variables to be interpreted as fields (functions,
or more generally sections) on the emergent $S^2$.

The coherent state satisfies a resolution of identity 
\be
\id_N =\frac{N}{4\pi} \int d \O \ket{\O}\bra{\O}.
\ee
Therefore
\be \label{F-int}
\Tr F = \frac{N}{4\pi} \int d \O F(\O),
\ee
where $F(\O) := \bra{\O} F \ket{\O}$ is the Berezin symbol of the matrix.
The normalization of \eq{F-int} can be checked with $F=1$. It is
illuminating to understand the formula \eq{F-int} in terms of
the coherent state representation. Note that the Hilbert
space of the spin-$J$ representation has dimension $N$. This divides
the fuzzy sphere into $N$ cells with a volume of
\be
\D \Omega_{\rm cells} = \frac{4 \pi}{N}
\ee
and the sum of a function $F(\O)$ over the cells
becomes an integral in the large $N$ limit
\be
\sum_i F(\Omega_i) = \frac{1}{\D \Omega_{\rm cells}} \int d\O F(\O).
\ee
This is \eq{F-int} in the coherent state representation. 

At finite $N$, the symbol of $ FG $ is not exactly
the pointwise product, but
defines a noncommutative star product
\cite{Grosse:1993uq,Presnajder:1999ky}
\be
\bra{\O} FG \ket{\O} = F(\O) * G (\O).
\ee
It has the exact finite expansion
\be
F*G = \sum_{l=0}^{2J}\frac{(2J-r)!}{r! (2J)!}
K_{a_1 b_1} \cdots K_{a_l b_l} \del_{a_1} \cdots \del_{a_l} F
\del_{b_1} \cdots \del_{b_l} G  , \qquad K_{ab} = P_{ab} + i \o_{ab}, 
\ee
where $P_{ab}$ is the tangential projector and
$\o_{ab}$ is the sphere symplectic tensor,
\be
P_{ab} = \d_{ab} -n_a n_b, \qquad
\o_{ab}=\e_{abc} n_c.
\ee
For smooth configurations,
\be
 F* G  = FG + \frac{1}{2J} K_{ab} \del_a F \del_b G + \cdots 
\ee
In intrinsic coordinates,
$K_{ab} \del_a F \del_b G = D_A F D^A G + i \e^{AB} D_A F D_B G$. This
gives rise to the star-commutator
\be
[F,G]_* = \frac{i}{J}\{F, G\}_{S^2} + O(J^{-2}),
\ee
where
\be
\{F, G\}_{S^2} = \e^{AB} D_A F D_B G = \e_{abc} n_a \del_b F \del_c G
\ee
is the  Poisson bracket on $S^2$. For smooth configurations,
\be
[F,G]_* = O(N^{-1})
\ee
and so the large-$N$ continuum theory becomes Abelian.
The background commutator is different:
\be
[J_a, F] \to -iK_a{}^A D_A F
\ee
remains finite at large $N$ since $J \sim N$ in the background compensate
the $1/N$ noncommutative. Therefore $[J_a, F]$ becomes an ordinary spatial
derivative, whereas $[F,G]$ between two smooth fluctuations is suppressed.

In general, a
$N \times N$ matrix can be expanded in terms of the fuzzy spherical
harmonics
$\Yh_{lm}$ 
\be
F = \sum_{l =0}^{2J} \sum_m F_{lm} \Yh_{lm}.
\ee
This gives the Berezin symbol
\be
F(\O) = \sum_{l,m} F_{lm} \Yh_{lm}(\O),
\ee
where $\Yh_{lm}(\O) \simeq Y_{lm}(\O)$, the usual spherical harmonics
when $l \ll J$. The sum is restricted to $l \leq 2J = N-1$
and is due to the
angular cutoff of the fuzzy sphere. Hence the fuzzy sphere provides a
natural
rotational invariant regularization for QFT on the ordinary sphere
\cite{Grosse:1995ar,Chu:2001xi}.

\section{Berry Monopole on Fuzzy Sphere}
In
this appendix, we demonstrate another novel property of the fuzzy
sphere: that the fuzzy sphere carries  an intrinsic magnetic monopole
with charge proportional to the rank of the fuzzy sphere.

To see this, we note that since $g(\O)$ and $g(\O)e^{i\a(\O) J_z}$
both represent a rotation of the north pole to the same point $\O$, this
gives rise to a phase ambiguity in the definition of the coherent state:
\be \label{gt-Omega}
\ket{\O} \to e^{i J \a(\O)}\ket{\O}
\ee
and hence the coherent states form  a section of a nontrivial line bundle
over $S^2$. The associated Berry connection is
\be \label{A-polar}
\cA = -i \bra{\O} d \ket{\O} = J(1-\cos \th) d\phi,
\ee
and has the  curvature 
\be \label{BF}
\cF = d \cA = J \sin \th d \th d\phi.
\ee
In  terms
of the vielbein $E^1  = R d \th, E^2 = R \sin \th d \phi$, the field strength
$\cF  = \frac{J}{R^2} E^1 E^2 $ is constant.  
This is a monopole with the first Chern number
\be
c_1 = \frac{1}{2\pi} \int_{S^2} \cF = 2J = N-1. 
\ee
We note that the Berry monopole  is
not an arbitrary external magnetic field
and does not  correspond to any propagating degrees of freedom.
It is topological in origin and its charge is conserved
due to a topological conservation law.
In \cite{Chu:2026wyh},
it was proposed that tunneling of fuzzy sphere gives
the fundamental quantum mechanical description for the black hole
%c11 Hawking
radiation process of quantum black hole. Now, with the monopole background
explicitly established on the fuzzy sphere, we can understand the origin
of the small charge monopole that was created in the tunneling process
there:
In fact as a
fuzzy sphere tunnels to a small fuzzy sphere,
it's rank decrease and a small charge monopole must be created due to
conservation of monopole charge. 

We remark that the same spin-$J$ structure \eq{JJ} of the
fuzzy sphere determines both the
fuzzy noncommutativity and the Berry monopole flux.
In fact, 
the Berry monopole and the noncommutative geometry
are simply two descriptions of the same quantization of $S^2$.
To see this, we start from the monopole and  obtain a symplectic two-form
on the sphere,
\be
\o_J = \cF = J \sin \th d\th d\phi.
\ee
The inverse of which define a Poisson bracket
\be
\{f,g\}_{\o_J} = \frac{1}{J\sin \th} (\del_\th f \del_\phi g - \del_\phi f \del_\th g).
\ee
For the coordinate functions $n_a$ on the fuzzy sphere,
we have
\be
\{n_a, n_b\}_{\o_J} = \frac{1}{J}\e_{abc} n_c.
\ee
Quantization gives precisely  the fuzzy noncommutative geometry
\be
[n_a, n_b] =  \frac{i}{J}\e_{abc} n_c.
\ee 

It is interesting to note that
the monopole is invisible to adjoint matrices and does not
play any role in matrix models such as BFSS or IKKT models before.
On the other hand, it is feelable to fermions in the fundamental
representation, such as those in our model.

\section{Multiple $\KK \geq 3$ interacting blocks}

Consider a system of $\KK \geq 3$ blocks with the node ranks given
by $N_i$, $i=1,\ldots,\KK$,
let us construct the system matrix
variables $\mathbf X_a$ and $\mathbf\Psi$ defined
by
\be
(\mathbf X_a)^{ii}=X_a^{(i)},
\qquad
(\mathbf X_a)^{ij}=W_a^{ij}=\sqrt{M_0}\,Z_a^{ij}
\quad(i\ne j),
\ee
with $Z_a^{ji}=Z_a^{ij\dagger}$, and
\be
(\mathbf\Psi)^{ii}=\psi_i,
\qquad
(\mathbf\Psi)^{ij}=\eta^{ij}
\quad(i\ne j),
\ee
where the oriented Grassmannian links $\eta^{ij}$ and $\eta^{ji}$ are independent.
The square matrices $\mathbf X_a$ and $\mathbf\Psi$ are of dimension
\be
N_{\rm tot}=\sum_{i=1}^{\KK}N_i.
\ee

With
\be
\widehat\lambda :=\sum_i\lambda_iP_i,
\qquad
\lambda_i:=\frac{\MP}{N_i^2},
\ee
where $P_i$ are the projector to the $i^{\rm th}$ diagonal block,
we propose to consider the following matrix interacting Lagrangian for the
system
\be
L_{\KK}=
\frac{1}{2M_0}\Tr\dot{\mathbf X}^2_a
+
\Tr\!\left[
\widehat\lambda
\left([\mathbf X_a,\mathbf X_b]^2+4\mathbf X_a\mathbf X_a\right)
\right] +
i\Tr\dot{\mathbf\Psi}^{\dagger}\mathbf\Psi
- a_2\Tr\!\left(
\widehat\lambda\,
\mathbf\Psi^{\dagger}\sigma^a\mathbf X_a\mathbf\Psi \right) .
\label{compactK}
\ee
At vanishing off-diagonal bosonic and fermionic link fields,
\be
L_{\KK}
=
\sum_{i=1}^{\KK}L^{(N_i)}[X^{(i)},\psi_i].
\ee
On the block configurations, the theory has a residual
symmetry $\prod_iU(N_i)/U(1)$
after imposing the trace condition.

To display the interactions, let us introduce
\bea
&& F_{ab}^{(i)}:=[X_a^{(i)},X_b^{(i)}], \\
&& h_{ab}^{ij} :=\mathcal D_a^{ij}Z_b^{ij}-\mathcal D_b^{ij}Z_a^{ij},
\qquad \mbox{where
$D_a^{ij}Z_b^{ij} :=X_a^{(i)}Z_b^{ij}-Z_b^{ij}X_a^{(j)}$},\\
&& q_{ab}^{(i)} := \sum_{j\ne i} \left( Z_a^{ij}Z_b^{ji}-Z_b^{ij}Z_a^{ji} \right),
\qquad 
s_{ab}^{ij} := \sum_{k\ne i,j}
\left( Z_a^{ik}Z_b^{kj}-Z_b^{ik}Z_a^{kj} \right)
\quad \mbox{for $i\ne j$},
\label{sK}
\eea
The commutator blocks are
\begin{align}
[\mathbf X_a,\mathbf X_b]^{ii}
&=F_{ab}^{(i)}+M_0 q_{ab}^{(i)},\\
[\mathbf X_a,\mathbf X_b]^{ij}
&=\sqrt{M_0}\,h_{ab}^{ij}+M_0 s_{ab}^{ij}
\qquad(i\ne j).
\end{align}
The canonically normalized smooth bosonic link action is
\be
L^{(\KK)}_{B,\mathrm{link}}
=
L^{(\KK)}_{B,\mathrm{kin}}
+
L^{(\KK,2)}_{B}
+
L^{(\KK,3)}_{B}
+
L^{(\KK,4)}_{B},
\ee
where
\bea
&& L^{(\KK)}_{B,\mathrm{kin}}
= \sum_{i<j}\Tr\dot Z_a^{ij\dagger}\dot Z_a^{ij}+
4M_0\sum_{i<j}(\lambda_i+\lambda_j) \Tr Z_a^{ij\dagger}Z_a^{ij},
\label{massKexact} \\
&& L^{(\KK,2)}_{B}=
-M_0\sum_{i<j}(\lambda_i+\lambda_j)
\Tr(h_{ab}^{ij}h_{ab}^{ij\dagger}) + 2M_0\sum_i\lambda_i
\Tr_{N_i}\!\left(F_{ab}^{(i)}q_{ab}^{(i)}\right), 
\label{quadraticKexact}\\
&& L^{(\KK,3)}_{B} =
-M_0^{3/2}\sum_{i<j}(\lambda_i+\lambda_j)
\Tr\!\left(
h_{ab}^{ij}s_{ab}^{ij\dagger}
+s_{ab}^{ij}h_{ab}^{ij\dagger}
\right),
\label{cubicKexact} \\
&& L^{(\KK,4)}_{B}=
M_0^2\sum_i\lambda_i
\Tr_{N_i}\!\left(q_{ab}^{(i)}\right)^2-
M_0^2\sum_{i<j}(\lambda_i+\lambda_j)
\Tr(s_{ab}^{ij}s_{ab}^{ij\dagger}).
\label{quarticKexact}
\eea
Here $L^{(\KK,n)}_{B}$ arises from the YM-commutator term of the matrix model,
and $n= 2,3,4$ denotes the order of link variables $Z$. 
We note that
for $\KK=2$, $s_{ab}^{ij}=0$, so the cubic term ($n=3$)
vanishes and the formulas
reduce to \eqref{boslink2}.  For $\KK\ge3$, the cubic terms are 
triangular 3-bodies
interactions and the quartic terms are closed-path interactions among links.
The generic difference between 2 bodies and 3 bodies
interaction is a character of gravitational interaction beyond the
Newtonian level.

For completeness, we also give the interacting fermionic action,
It is useful to use a unified notation
\be
\Psi^{ij}:=(\mathbf\Psi)^{ij}
=
\begin{cases}
\psi_i,&i=j,\\
\eta^{ij},&i\ne j.
\end{cases}
\ee
Then the full fermionic part of \eqref{compactK} is
\be
\begin{aligned}
L_F^{(\KK)}=
i\sum_{i,j}
\Tr_{N_j}\!\left(\dot\Psi^{ij\dagger}\Psi^{ij}\right)
- a_2\sum_j\lambda_j
\sum_{i,k}
\Tr_{N_j}\!\left(
\Psi^{ij\dagger}\sigma^a
(\mathbf X_a)^{ik}
\Psi^{kj}
\right).
\end{aligned}
\label{fermionKcompact}
\ee
The dimensions in every term are
\be
\Psi^{ij\dagger}:N_j\times N_i,
\qquad
(\mathbf X_a)^{ik}:N_i\times N_k,
\qquad
\Psi^{kj}:N_k\times N_j.
\ee
so the trace is over the right endpoint $N_j$.
Consequently the coefficient selected by the projector prescription is exactly
$ y_j=a_2\lambda_j$.
Subtracting the diagonal fermion action gives explicitly, for the
fermionic link sector
\be
\begin{aligned}
L^{(\KK)}_{F,\mathrm{link}}={}&
i\sum_{i\ne j}
\Tr_{N_j}\!\left(\dot\eta^{ij\dagger}\eta^{ij}\right) -
a_2\sum_j\lambda_j
\sum_{i\ne j}
\Tr_{N_j}\!\left(
\eta^{ij\dagger}\sigma^aX_a^{(i)}\eta^{ij}
\right)
\\
&-
a_2\sqrt{M_0}
\sum_j\lambda_j
\sum_{i\ne k}
\Tr_{N_j}\!\left(
\Psi^{ij\dagger}\sigma^aZ_a^{ik}\Psi^{kj}
\right).
\end{aligned}
\label{fermlinkK}
\ee
The last line contains all Yukawa interactions involving a bosonic link.
Depending on the indices, it includes the processes:
(a) conversion between a node fermion and an oriented fermionic link;
(b) interactions between two fermionic links and one bosonic link;
(c) for $\KK\ge 3$, fermionic triangle-path interactions.

\noindent\underline{FIDO frame}

For many-body interaction, e.g. single black hole with interior
and exterior blocks, or multiple black holes with environment,
one can fix attention to one of the black hole blocks and use its
FIDO time to describe the physics. Let's say we use the FIDO time
of the $i =1$ ($N_1 =N$) block
\be
  \tau= \frac{2a_0 M_P l_P}{N}\,t.
\ee
Correspondingly, define for each block
\be
  \alpha_i:=\frac{N^2}{2N_i^2 l_P^2},
  \qquad
  \widehat\alpha:=\sum_{i=1}^\cK \alpha_i P_i .
  \label{eq:alpha-def}
\ee
In this case, the action for each block is
\be
  S_i =\frac{l_P}{a_0N}\int d\tau 
  \Tr_{N_i}\left[
    \dot X_a^{(i) 2}
    +\alpha_i ([X_a^{(i)},X_b^{(i)}]^2+4X_a^{(i)2})
  \right]+\int d\tau\,
 \Tr_{N_i} \left[
    i\psi_i^\dagger\dot\psi_i
    -{a_2 \frac{l_P}{a_0N}}\alpha_i
   \psi_i^\dagger\sigma^aX_a^{(i)}\psi_i
  \right] . \nn
 \ee
 To describe the link interaction, let us introduce
 \be
  F_{ab}^{(i)}:=[X_a^{(i)},X_b^{(i)}] .
  \ee
  \bea
  && h_{ab}^{ij}:=D_a^{ij}W_b^{ij}-D_b^{ij}W_a^{ij} , \quad
  \mbox{where}\quad
  D_a^{ij}W_b^{ij}:=X_a^{(i)}W_b^{ij}-W_b^{ij}X_a^{(j)}, \\
  &&  q_{ab}^{(i)}:=
   \sum_{j\neq i}\left(W_a^{ij}W_b^{ji}-W_b^{ij}W_a^{ji}\right), \quad
    s_{ab}^{ij}:=
    \sum_{k\neq i,j}\left(W_a^{ik}W_b^{kj}-W_b^{ik}W_a^{kj}\right) \;
    \mbox{for $i \neq j$}.
  \eea
And the bosonic link interaction is
\be
S_{B, {\rm link}}^{(\cK)}=\int\dd\tau\,L_{B, {\rm link}}^{\cK},\qquad
L_{B, {\rm link}}^{(\cK)}
  =L_{B,\mathrm{kin}}^{(\cK)}+L_B^{(\cK,2)}+L_B^{(\cK,3)}+L_B^{(\cK,4)},
  \ee
where
\bea
&&  L_{B,\mathrm{kin}}^{(\cK)}
  ={ \frac{l_P}{a_0N}}
  \left[
  2\sum_{i<j}\Tr_{N_i}\!\left(\dot W_a^{ij}\dot W_a^{ji}\right)
  +4\sum_{i<j}(\alpha_i+\alpha_j)
  \Tr_{N_i}\!\left(W_a^{ij}W_a^{ji}\right)
  \right] , \\
&&  L_B^{(\cK,2)}
  =\frac{l_P}{a_0N}
  \left[
  -\sum_{i<j}(\alpha_i+\alpha_j)
  \Tr_{N_i}\!\left(h_{ab}^{ij}h_{ab}^{ij\dagger}\right)
  +2\sum_i\alpha_i\Tr_{N_i}\!\left(F_{ab}^{(i)}q_{ab}^{(i)}\right)
  \right] , \\
&&  L_B^{(\cK,3)}
  =- \frac{l_P}{a_0N}
  \sum_{i<j}(\alpha_i+\alpha_j)
  \Tr_{N_i}\!\left(
    h_{ab}^{ij}s_{ab}^{ij\dagger}
    +s_{ab}^{ij}h_{ab}^{ij\dagger}
  \right) , \\
&&  L_B^{(\cK,4)}
  = \frac{l_P}{a_0N}
  \left[
  \sum_i\alpha_i\Tr_{N_i}\!\left(q_{ab}^{(i)}q_{ab}^{(i)}\right)
  -\sum_{i<j}(\alpha_i+\alpha_j)
  \Tr_{N_i}\!\left(s_{ab}^{ij}s_{ab}^{ij\dagger}\right)
  \right] .
\eea
As above, for $\cK\geq3$, $L_B^{(\cK,3)}$ is a triangular three-node
interaction, while the $s_{ab}^{ij}s_{ab}^{ij\dagger}$ term in $L_B^{(\cK,4)}$
is a closed-path quartic interaction involving more than one link.
Similarly, the action for the fermionic link sector reads
\bea
  L_{F,{\rm link}}^{(K)}
  =&\;i\sum_{i\neq j}\Tr_{N_j}\!\left(\eta^{ij\dagger}\dot\eta^{ij}\right)
  -{a_2 \frac{l_P}{a_0N}}
  \sum_j\alpha_j\sum_{i\neq j}
  \Tr_{N_j}\!\left(
    \eta^{ij\dagger}\sigma^aX_a^{(i)}\eta^{ij}
  \right)
  \\
  &-{a_2 \frac{l_P}{a_0N}}
  \sum_j\alpha_j\sum_{i\neq k}
  \Tr_{N_j}\!\left(
    \Psi^{ij\dagger}\sigma^aW_a^{ik}\Psi^{kj}
  \right) .
\eea

\section*{Acknowledgments}

We thank Sumit Das, Dimitrios Giataganas, Pei-Ming Ho, Norihiro Iizuka,
Satoshi Iso, Asuka Ito, Hikaru Kawai, Hsiang-Nan Li,
Ian Low, Emil Martinec, Samir Mathur,
Jiro Soda, Harold Steinacker for discussions and comments.
We
acknowledge the support of this work by NCTS, the National Science and
Technology Council of Taiwan for the grant 113-2112-M-007-039-MY3, and
the National Tsing Hua University 2025 Talent Development Fund for a
TSAI WANG, YUAN-YANG Distinguished Talent Chair Professorship.

\bibliographystyle{utphys}

\bibliography{references}  
\end{document}